\documentclass[a4paper,11pt]{article}
\usepackage{latexsym}
\usepackage{amsmath}
\usepackage{amsfonts}
\usepackage{amssymb}
\usepackage{graphicx}
\usepackage[latin1]{inputenc}
\usepackage{color}
\usepackage{float}
\usepackage{setspace}
\usepackage{authblk}

\begin{document}

\title{Imitation, Proximity, and Growth \\
\textit{\Large{A Collective Swarm Dynamics Approach}}}

\author[1]{Olivier Gallay\thanks{\texttt{olivier.gallay@unil.ch; postal address: University of Lausanne, Faculty of Business and Economics (HEC Lausanne), Department of Operations, Quartier UNIL-Chamberonne, CH-1015 Lausanne, Switzerland (corresponding author)}}}
\author[2]{Fariba Hashemi\thanks{\texttt{fariba.hashemi@ki.se}}}
\author[3]{Max-Olivier Hongler\thanks{\texttt{max.hongler@epfl.ch}}}
\affil[1]{\small{University of Lausanne, Switzerland}}
\affil[2]{\small{ETH Zurich (ETHZ), Switzerland, and Karolinska Institutet, Sweden}}
\affil[3]{\small{EPFL, Switzerland}}
\date{}
\maketitle



\abstract{
\vspace{0.2cm}

\noindent This paper is based on the premise that economic growth is driven by an interplay between innovation and imitation in an economy composed of interacting firms operating in a stochastic environment. A novel approach to modeling imitation is presented, based on range-dependent processes that describe how firms consider proximity when imitating peers who are found in a given neighborhood in terms of productivity. Using a particularly tractable approach, we are able to analyze how drastically different economic growth scenarios emerge from different imitation strategies. These emerging scenarios range from diffusive growth where the variance of productivity grows indefinitely, to balanced growth described by a traveling wave with fixed variance. The latter scenario is sustained only when imitation strength among firms exceeds a critical bifurcation threshold.

\vspace{0.5cm}


\vspace{0.5cm}
\noindent
}


\newpage
\section{Introduction}

Economic growth results from a complex interplay between a plurality of factors and among these, the firms' productivity innovation activity and/or imitation mechanism of actual technological leaders are strongly determinant (\textit{e.g.} \cite{IWAI2000, LUCAS2009,LUTTMER2012a}). Accordingly, any progress towards a refined understanding of how the innovation and imitation dynamics operate and coexist, brings us closer to fine-tune (and will ultimately allow to optimize) one of the key factors underlying economic growth. This essential objective has motivated mathematical modeling efforts in the economics literature for decades, aiming to isolate a possibly restricted number of parameters directly relevant for the growth process (\textit{e.g.} \cite{STALEY2011, KOENIG2016, BENHABIB2017}). The present paper falls within this general scope by unveiling a class of exactly solvable multi-agent dynamics for which the interplay between innovation and imitation can be analytically discussed. The log-productivity dynamics of firms (interchangeably referred to as agents in this paper) co-evolving in an economy is modeled with the help of interacting nonlinear random walkers evolving in discrete time on discrete productivity echelons. Our approach offers the following three-fold contribution \textit{i)} it provides, in the state-space-and-time continuous limit, a new class of nonlinear partial differential equations which can be analytically discussed,  \textit{ii)} it connects economic growth with the imitation strength between the firms and \textit{iii)} it unveils a bifurcating  transition from a diffusive to a propagating growth regime, which is tuned by the relative influence played by the firms with productivity close to the technological frontier.\\

\noindent The present modeling framework originates from \cite{LUCAS2009}\footnote{The same vision is later adopted by \textit{e.g.} \cite{RIDLEY2010, STALEY2011, LUTTMER2012a, LUCAS2014}.}, who considers a growing economy resulting from the emergence of a collective dynamic pattern generated by a large swarm of mutually interacting (and possibly stochastic) agents. The improvement in productivity achieved by each firm ultimately generates economic growth. Specifically,  productivity growth is  understood to be driven by the joint action of \textit{(i)} a sustained flow of innovative attempts that are subject to random fluctuations and \textit{(ii)} an imitation process  among the agents that acts as a rectifying mechanism, thereby ensuring that only productive ideas are ultimately retained. 
This highlights that a growing economy has to always be sustained out of equilibrium \cite{IWAI1984},
as there is a constant need for technology leaders who, via innovation attempts that always come along with noise, generate novel ideas and processes. 
The innovative breakthroughs are then tested and evaluated, and ultimately, only the best ones are adopted by concurrents. This dynamic co-action between innovation and imitation processes is a key factor in sustaining economic growth. Imitation of leading peers filters out the poor results inherent to any risky innovation moves; it steadily scavenges positive outputs from the intrinsically noisy innovative environment. Economic growth results therefore from a subtle interplay between a fluctuating mechanism (innovation) and a mechanism that is deterministic (imitation). Eliminating either of these two basic mechanisms leads to a reduction in (or even a cancellation of) progress, as it is also expressed 
in \cite{REINGANUM1985}.\\

\noindent To proceed towards a mathematical stylization of the above ideas, as  in \cite{LUCAS2009, LUTTMER2012a, LUCAS2014, BENHABIB2017}, we consider a collection of firms each of which is described by a Markov chain (MC) evolving on a scalar ladder of log-productivity echelon. While in \cite{LUCAS2009, LUTTMER2012a, LUCAS2014, BENHABIB2017}, several MC echelons can be crossed during a single time step, in this paper we focus on MCs with jumps confined to the nearest echelon only, \textit{i.e.} birth-and-death (BD) processes. The imitation mechanism is constructed by assuming the MC's birth rate to be monotonously increasing with the imitation intensity in the economy. Specifically, each firm gathers in real-time the productivity level of competitors drawn from a representative sample, and adjusts its imitation rate accordingly (\textit{i.e.} the MC's birth rate). As in  \cite{LUCAS2009, LUTTMER2012a, LUCAS2014, BENHABIB2017}, we implement an aversion for being a productivity laggard by imposing  each firms' birth rate to be proportional to the (time-dependent) number of observed leaders. Such an imitation rule implements a nonlinearity into the MC's dynamics which usually precludes analytical studies of the resulting transient behavior. However, when examined in a continuous state-space-and-time limit, our modeling framework illustrates that the log-productivity dynamics of a representative firm can be  effectively described by a nonlinear diffusion process of the form: 
\begin{equation}
\label{MFAPP}
\left\{
\begin{array}{l}
{d X(t)} =\left[\,  \, \alpha  + {\cal J}(X(t), \rho(x,t))\,  \, \right]  dt + \sigma dW(t),   \\ \\ 
\int_{\mathbb{R} }\rho(x,t) dx \equiv 1.
\end{array}\right.
\end{equation}

\noindent In Eq.(\ref{MFAPP}), $X(t)$ stands for  log-productivity of a representative firm in a continuous state-space-and-time limit, the normalized quantity $\rho(x,t)$ stands for the log-productivity distribution of the firms in the economy (in other words, 
$\rho(x,t)dx$ represents the density of firms with log-productivity located in the interval $[x, x+dx]$ at time $t$), and the drift component ${\cal J}(X(t), \rho(x,t))$ describes the imitation mechanism. Note that ${\mathcal J}(X(t), \rho(x,t))$ is a positive definite function. The representative firm effectively interacts with the whole economy via the $\rho(x,t))$-dependency in ${\cal J}(X(t), \rho(x,t))$, which explicitly implements a nonlinear evolution. The constant drift intensity $\alpha \geq 0$ describes a systematic innovation propensity towards progress, and finally $dW_t$ stands for the standard White Gaussian Noise process (WGN) which models  the randomness of the  environment.\\

\noindent In this general context,the following questions will be addressed:


\vspace{0.1cm}
\begin{itemize}
\item[\textit{(a)}]  {\it Exogenous strategy and the sampling size effect.}\\
\noindent Imitation processes, or the attention that firms pay to their peers' productivity state, depends on information gathering or technology replication on either samples of the population or on the whole society. Do stable collective productivity evolutions emerge (\textit{i.e.} stationary balanced growth paths) for arbitrary mutual agent interactions? What is the importance of the sample size, and of the relative weight that firms attach to the observation of their peers (either firms close to their own productivity state, or productivity leaders)? Does there exist a critical imitation threshold that differentiates between different growth regimes?\\

\vspace{-0.1cm}
To address these issues, we allow the nonlinear imitation drift ${\cal J}(X(t), \rho(x,t))$ in Eq.(\ref{MFAPP}) to effectively depend on a parameter $U$ which measures the abstract distance separating the firms' log-productivities. More precisely, a firm with log-productivity $X(t)$ adjusts its imitation drift ${\cal J}\left(X(t), \rho(x,t)\right)$ by numbering its technological leaders located in a range $U$, namely those with log-productivities belonging to the interval $[X(t), X(t)+U]$. Accordingly, a large neighborhood $U$ implies that even far remote leaders are influential, thus describing strongly competitive environments. Conversely, when $U$ is small, firms are comparatively more cautious by being sensitive only to leaders close to their own log-productivity state. If imitation costs were implemented\footnote{By naturally assuming that imitation of a far remote productivity leader incurs a heavier cost compared to a close one.}, as actually done in \cite{KOENIG2016, BENHABIB2017}, small neighborhood $U$ sensitivities would reflect risk aversion. 
Not only the number of firms with log-productivities found within a neighborhood $U$, but also their relative distance from the productivity leaders, may influence, via ad-hoc weighting factors, the imitation propensity ${\cal J}(X(t), \rho(x,t))$. To highlight this aspect, we are able to unveil a bifurcation  separating two drastically different economic growth scenarios. When the level of mutual interactions  in the economy is below a critical threshold, a diffusive behavior dominates and growth cannot be sustained (\textit{i.e.} an evanescent propagating wave is generated). Conversely, when the imitation strength exceeds the bifurcation threshold, the diffusive noise effect due to the endogenous environment is counterbalanced by the firms' mimetic propensity, giving rise to a balanced growth path (\textit{i.e.} a propagating wave with constant variance). A similar bifurcation separating the two above propagation modes was also qualitatively illustrated by A. de Geus in \cite{DEGEUS2002}, where an example from ornithology was borrowed and applied to economics (a more detailed account is provided in Appendix B). Like what is happening here, a reduction in the agents' observation range decreases the importance of the mutual interactions, which ultimately precludes the possibility to generate stable collective scenarios (\textit{i.e.} for birds, the tendency to flock). This illustration confirms that below a critical threshold, mutual interactions are too weak to sustain a flocked evolution (corresponding here to a mirror image of  a balanced growth path), \cite{CUCKER2007, HONGLER2014}.

\vspace{0.2cm}
\item[\textit{(b)}]  {\it The role of the random environment.}\\
How does the randomness due to the firms' innovation attempts possibly add a hidden benefit to the overall growth process?
\vspace{0.2cm}
\item[\textit{(c)}] {\it Mean-field imitation game as an endogenous strategy.}\\
\noindent  Can we  interpret some of the observed emerging growth scenarios as resulting from an optimal strategy adopted in a multi-player game?  Is it possible to relate exogenous versus endogenous interaction rules leading ultimately to the same balanced growth path?\\

Expressed formally, the drift  $\left[\,  \, \alpha  + {\cal J}(X(t), \rho(x,t))\,  \, \right]$ should itself result from an optimal control  problem involving a large collection of players (\textit{i.e.} the firms), for which {\it mean-field games} (MFG) framework is naturally  suited. Specifically, we consider the following class of optimal control problems: 

\begin{equation}
\label{MFGBASE0}
\left\{
\begin{array}{l}
dX(t) = a^{*}[X(t), t] dt + \sigma dW(t), \\ \\
J\left(a(\cdot), X(\cdot)\right) =   \min\limits_{\left\{a \in \mathbb{A}\right\}} \mathbb{E} 
\left\{ 
 \int_{0}^{T}  \mathcal{L}(a(t), \rho(\cdot,s), X(t),t)  
 \right\} + c_T(X(T)),\\ \\ 
 \mathcal{L}\left(a(t), \rho(\cdot,s), X(s),s):=  c\left(a(t), X(t),t\right) - 
 V\left[\rho(\cdot,s),  X(s)\right], 
 \right\} 
\end{array}
\right.
\end{equation}

\noindent where the operator  $ \mathbb{E} \left\{\cdot \right\} $ is the expectation over the possible realizations of the noise source $dW(t)$, $T$ is a final fixed time horizon, ${\cal L}\left(x(s), \rho(\cdot,s) \right)$ is a running  cost function\footnote{The notation ${\cal L}\left(x(t), \rho(\cdot,t) ,t\right)$ means that the function ${\cal L}$ associates with the density function $\rho(\cdot,t)$ another function ${\cal L}\left(\rho(\cdot,t) \right)$ itself evaluated at $x$. Accordingly, a given  agent $x$ interacts with her peers only via their density function (in the mean-field approach, agents are indistinguishable).}, $\rho(x,t)$ is the density  of firms with log-productivity located in the interval $[x, x+dx]$ and $\mathbb{A}$ stands for the set of all admissible drifts $a[X(t), t] $ among  which the minimal $a^{*}[X(t), t] $ is to be found. In the sequel, we will make use of the framework given by Eq.(\ref{MFGBASE0}) and  explicitly calculate the  mean-field  interaction potential $V\left[\rho(\cdot,s),  X(s)\right]$ that gives rise to the propagating wave derived for the  exogenous rule evoked above in {\it (a)}. 

\end{itemize}

\noindent The possibility to bridge the gap between statistical physics (involving a large number of microscopic variables) and thermodynamics (involving a few macroscopic variables, i.e.,  those used in the real gas  equation of van de Waals) is a major success of theoretical physics. This achievement is inspiring and leads naturally to the question of whether a similar program could  be achieved for a "gas"  of economic agents. Clearly, a set of interacting gas particles is likely to behave in a far simpler way than a gas of  interacting "intelligent" agents who track an individual  goal, namely the maximization of a private utility function. In a  mathematically stylized way, MFGs offer one possibility to bridge the gap between the microscopic and macroscopic collective dynamics for large swarms of  such intelligent interacting agents. Our model fully belongs to the ongoing research activity oriented along this general line. Specifically, the new class of exactly solvable models presented in this paper shares many features with the celebrated class of Kuramoto coupled phase oscillators (KPOs) \cite{ACEBRON2005}. Like for the KPO dynamics, exogenous long-range mutual interactions produce a behavioral bifurcation from a desynchronized to a synchronized collective evolution. Similarly to the KPO dynamics,  a corresponding MFG can be constructed, thus enabling to unveil the corresponding endogenous rules obtained from individual utility function optimization \cite{YIN2012}. While for KPOs, the agents evolve on compact states (i.e., oscillator phases are confined on the circle), in our case the agents  evolve on the whole real line. This unwrapping of the state space is not an innocuous difference between our modelling framework and the KPO model. Indeed, a KPO-type bifurcation can only be obtained at the expense of introducing a barycentric weighting factor of the agent interactions. The  agent evolution on the linear state space offers the possibility to define ranks between the agents (i.e., laggards and leaders) and hence to complement the list of existing economic models based on the Schumpeterian quality ladder dynamics.


\subsection{Literature Review}

\noindent The economics literature is rich with theories of innovation and imitation dynamics which attempt to capture many realistic aspects of long-run economic growth.\\ 

\noindent In \cite{ROMER1990}, \cite{GROSSMAN1991}, and \cite{AGHION1992}, models of long-run growth, based on endogenous technological change in patent-protected environments, enable us to think about the determinants of technological progress and how the manner in which resources are allocated has differential impact on long-run productivity growth. This axis of research is referred to as endogenous growth theory and focuses on innovation-based growth dynamics. While \cite{ROMER1990} relies on models in which productivity growth is caused by innovative investments and the creation of new varieties of products, \cite{AGHION1992} address the Schumpeterian paradigm, following which innovation and creative destruction (i.e., when innovative technology tends to make older products obsolete) creates long-run economic growth. To that aim, \cite{AGHION1992} build their model on quality ladders, with respect to which an existing product can be substituted by a new innovative one.\\

\noindent The present paper can be seen as a natural generalization of the related Schumpeterian innovation-imitation dynamics initially introduced in \cite{IWAI1984}, and later extended in \cite{HENKIN1991, IWAI2000}. The common starting point for these studies is an evolution equation for the agent density $\rho(x,t)$ representing an abstract productivity level $x$ at a given time $t$. However, while in \cite{IWAI1984, HENKIN1991, IWAI2000}, each agent's drift is determined in real-time by the interactions that follow from an infinite observation range (\textit{i.e.} a given agent is influenced by all of her leaders or all of her laggards), we allow in the present paper the observation range to be an exogenously controlled variable (\textit{c.f.} Section \ref{sct_exo}). As stated above, this additional degree of freedom unveils a new, range-dependent, transition between two drastically different growth regimes. In \cite{IWAI1984, HENKIN1991, IWAI2000}, as the agent interactions are long-range, only stable and stationary balanced growth paths are observed.\\

\newpage
\noindent Building on the paradigm under which economic ideas can be ranked according to their productive usefulness on the rungs of a scalar quality ladder, an agent-based model is proposed in \cite{STALEY2011} to study some characteristics of economic growth. Economic agents have an incentive to adopt a higher productive state by jumping at random times either to a higher rung that is already occupied by another agent (the imitation process) or to a higher rung without any side considerations (the innovation process). For a large population of agents and for ladders with a large number of rungs, the natural approach that is adopted is to describe the aggregated state of the agent swarm by a measure density function $\rho(x,t)$ that quantifies the density of agents  at a given position $x$ at a given time $t$. The approach derived in \cite{STALEY2011} illustrates how for a large population of agents, $\rho(x,t)$ solves a deterministic nonlinear reaction-diffusion equation that is drastically different than the Burgers' equation we will be dealing with in the present work. Indeed, the density $\rho(x,t)$ in \cite{STALEY2011} always exhibits a  stable traveling wave character (\textit{i.e.} a soliton), which represents a steadily growing economy.\\

\noindent \cite{LUTTMER2012a} and \cite{LUTTMER2012b} consider a competitive economy with entry and exit, and focus on describing the conditions that have to be fulfilled by new entrants for growth to be sustained. The randomness which comes along with the firms' innovation attempts is also explicitly taken into account, and it is clearly  emphasized that the  joint role played by innovation and imitation is essential to ultimately enable a balanced growth path. While imitation is shown to be a mandatory ingredient for the emergence of a stationary agent distribution, the specific class of  models constructed in \cite{LUTTMER2012a, LUTTMER2012b} highlight that even a relatively low imitation strength is already able to produce a balanced growth path where entry and exit rates are high.\\

\noindent In \cite{BENHABIB2017} (see also  \cite{BENHABIB2014}), an innovation-imitation dynamics is constructed, for which the innovation states are also stylized by positions on a ladder. The jumps on this  ladder are intermittently driven by an alternating innovation productivity state,  and the alternations themselves are governed by a continuous time two-state Markov chain.  This type of random environment ultimately implies the technological frontier to progress at finite velocity. This has to be contrasted with our present paper, where the  use of White Gaussian Noise (and the unbounded realizations thereof) leads to infinite frontier velocities. In addition, the imitation process, referred to as adoption in \cite{BENHABIB2017}, depends on a maximization process which is defined via an ad-hoc utility function. Hence, the modeling framework considered in \cite{BENHABIB2017} exhibits a higher degree of complexity since it includes an additional optimization step,  which ultimately precludes the possibility to derive an exact transient analysis.\\

\noindent The model introduced in \cite{LUCAS2014} considers a collection of agents who divide their time between producing goods and interacting with productivity leaders to improve their own capabilities. As in the present study, the dynamics is driven by an underlying stochastic environment, and a mean-field approach in continuous-time is adopted. The modeling approach adopted in \cite{LUCAS2014} relies on individual utility functions and focuses on the resulting stochastic optimal control problems (\textit{i.e.} one deals with an underlying MFG). In the context of an MFG, all players in the society are mutually interacting, and thus the observation range between the agents is effectively infinite. However, contrary to the class of individual objective functions that are discussed in Section \ref{sct_endo} of the present paper, a flocking-diffusive bifurcation does not exist in \cite{LUCAS2014}, even when the imitation intensity decreases with agent dispersion.\\

\noindent Similarly, productivity growth is modeled in \cite{KOENIG2016} as the outcome of two strategies, namely in-house research and development (R\&D; innovation) and replication of competitors' technology (imitation). Considering an infinite imitation range, the authors focus on the agents' choice between these two strategies, with individual profit maximization as the objective. It is shown that technology leaders tend to choose in-house R\&D as they get fewer imitation opportunities, while the cost-effective choice for technology laggards is to imitate more productive competitors.\\


\noindent The endogenous strategy that will be developed in Section \ref{sct_endo} relies on MFGs. Following the pioneering works of Lasry and Lions \cite{LASRY2006A, LASRY2006B, LASRY2007}, MFGs and their wide potential for applications have been triggering sustained interest in the economics literature (\textit{e.g.} \cite{LACHAPELLE2010, GUEANT2011, ACHDOU2014, GOMES2015, CARMONA2018}). For our particular class of models, we use a recent result exposed  in  \cite{SWIECICKI2016} to analytically derive the stationary productivity waves that correspond exactly to the MFG ergodic states.

\subsection{Outlook}

The rest of the paper is organized as follows. In Section \ref{sct_exo}, the log-productivity of an economy consisting of homogeneous interacting firms is modeled via a set of coupled discrete-time and discrete-space Markov chains. By taking the continuous state-space-and-time limit and adopting a mean-field approach, our nominal dynamics can be reduced to a nonlinear and nonlocal partial differential equation in $1+1$ dimensions (\textit{i.e.} one dimension for space (productivity) and one for time), which corresponds to the Chapman-Kolmogorov equation governing a strongly nonlinear diffusive process. For specific limiting regimes, characterized either by infinitesimal versus infinitely long imitation ranges in the log-productivity state space, the dynamics is observed to converge towards the Burgers' equation for which one is able to derive exact transient solutions. Two drastically propagating growth modes are explicitly unveiled, namely a diffusive versus a stable propagating wave. A slight generalization of the nominal model is then discussed in Section \ref{sct_mod}, where the influence of a leader is weighted by its log-productivity distance relative to the average of the entire firm population within the economy. Depending on this weight, which in the sequel is controlled by a single parameter, we are able to analytically  characterize the critical bifurcation point that separates the diffusive versus the collective stable propagating wave. In complement  to  the exact results, intermediate parameter ranges for which analytical results cannot be worked out are reported in a series of simulation experiments that are exposed in Appendix E. Section \ref{sct_endo} is devoted to the endogenous MFG approach where we are able to explicitly construct the mean-field potential, which after dynamic programming, gives rise to the identical collective propagating wave as the one derived in Section \ref{sct_exo}. Concluding remarks can be found in Section \ref{sct_conc}.

\section{Discrete Modeling of Innovation-Imitation  Dynamics}\label{sct_exo}

\noindent To model the innovation-imitation collective behavior of a collection of $N$ firms ${\cal A}_k$ with  $k=1,2,\dots, N$, we follow the lines exposed in  \cite{KOENIG2016} and consider a collection of $N$ scalar stochastic processes $X_k(t)$  which describe the  instantaneous  log-productivity states of the firms. As in \cite{KOENIG2016}, we assume that $X_k(t) \in \mathbb{Z}_a:= \left\{\cdots -2a, -a, 0, +a, +2a, \dots  \right\}$. Hence, the log-productivity states are described by a regular productivity ladder with echelon spacing $a$. We first consider a discrete time evolution  with  time-steps $\Delta$, and we write $X_k(t):= X_k(\nu \Delta )$, with $\nu \in \mathbb{N}^{+}$. Again along the lines drawn in \cite{KOENIG2016},  we assume that the ${\cal A}_k$ evolution can be stylized by a MC dynamics in which the jump transition probabilities  jointly depend on the underlying innovation and imitation processes. Specifically, innovation induces an  effective  positive average, denoted $\alpha_k \geq 0$,  and for the imitation the associated drift reads as ${\cal D}_k\left[\vec{X}(t)\right]>0 $, where  $\vec{X}(t):= \left(X_1(t), X_2(t), \cdots, X_N(t)\right)$. We emphasize that, while the innovation process is assumed to yield a constant drift component, the imitation component depends on the instantaneous productivity states occupied by the ${\cal A}_k$ concurrent fellows, as it is reflected by the informal notation ${\cal D}_k\left[\vec{X}(t)\right]$. From now on, we limit the discussion to a population of homogeneous firms\footnote{The same assumption is also implemented in \cite{KOENIG2016}.}, leading us to write: 
\begin{equation}
\left\{
\begin{array}{l}
\alpha_k = \alpha, \\ \\
{\cal D}_k\left[\vec{X}(\nu \Delta)\right]={\cal D}\left[\vec{X}(\nu \Delta)\right].
\end{array}
\right.
\end{equation}

\noindent The log-productivity dynamics of the firms is now stylized by a set of $N$ interacting nearest neighbor ladder rungs\footnote{These processes are also known as birth-and-death processes in probability theory.} in which the bias in the probability jumps effectively models the  innovation and imitation mechanisms. The mutual interactions in our collection of BD processes are due to an  imitation mechanism which is  implemented by  an {\it "avoid to be a log-productivity laggard"} rule:

\begin{itemize}
 \item[\textit{(a)}] For any  $k=1,2,\cdots, N$, agent ${\cal A}_k$ steadily  counts  the number ${\cal N}_k(t)$ of her log-productivity leaders\footnote{It is not strictly necessary to observe all other agents, as it would be actually sufficient to consider a representative statistics of the agent society in order to allow for the use of the mean-field approach.} ${\cal A}_j$ for  $j\neq k$, $j=1,2,\cdots, N$, this within an observation  range $(U\, a) \geq 0$, and hence ${\cal N}_k(t) = \sum_{j\neq k} \mathbb{I}\left\{ 0 \leq  \left[X_j(t) - X_k(t)\right] \leq U\right\}$\footnote{The function $\mathbb{I}\left\{ z\right\}$ is the indicator function which takes the value $1$ when $z$ is true and $0$ otherwise.}.

\vspace{0.2cm}
\item[\textit{(b)}] The $X_k(t)$ jump process is endowed with a positive definite  bias monotonously increasing with ${\cal N}_k(t)$.
\end{itemize}

\noindent Thanks to the homogeneity assumption and for large $N$, we adopt a mean-field approach (\textit{e.g.} \cite{HOPENHAYN1992}, \cite{LUTTMER2007}), and hence  focus on the evolution of a single  randomly selected firm ${\cal A}$  whose behavior will  be representative of the whole population. Within this picture, ${\cal N}(t) := {\cal N}(\nu \Delta)$ stands for the number of log-productivity leaders numbered by ${\cal A}$ within her observation range $U \, a$. Defining $P\left[(ka, \nu\Delta \right]$ to be the probability of finding ${\cal A}$ at position $ka$ at time $\nu \Delta$, the random evolution is formally described by the nonlinear BD master equation:
\small{
\begin{equation}
\label{FORMALDYN}
\left\{
\begin{array}{l}
 P\left[(ka, (\nu+1)\Delta \right]  =  p(ka, \nu \Delta, \vec{X}(\nu\Delta)) P\left[(k-1) a , \nu \Delta\right]  + q(ka, \nu \Delta,  \vec{X}(\nu\Delta)) P\left[(k+1) a , \nu \Delta \right] , \\ \\ 
 p(ka, \nu \Delta,  \vec{X}(\nu\Delta)) :=  {1 \over 2} \left\{ (1+ \alpha  )+ {\cal D}\left[ka, \ \nu \Delta, {\cal N}(\nu \Delta)\right] \right\},\\ \\ 

q(ka, \nu \Delta,  \vec{X}(\nu\Delta)) :=  {1 \over 2}  \left\{ (1 - \alpha) - {\cal D}\left[ka,  \nu \Delta, {\cal N}(\nu \Delta)\right] \right\},\\ \\

{\cal D}\left[ka, \nu \Delta, {\cal N}(\nu \Delta)\right] := \beta \left\{ \sum_{m= k}^{m = k+U} \left[ P(ma, \nu \tau)\right] \right\}  \in [0,1],
 \end{array}\right.
\end{equation}
}
\noindent where the exogenous parameters  $\alpha\geq 0$ and $\beta\geq 0$ are chosen in an ad-hoc range ensuring that  both BD jump probabilities $p(ka, \nu \Delta,  \vec{X}(\nu\Delta))$ and  $q(ka, \nu \Delta,  \vec{X}(\nu\Delta))$  remain positively defined. At this stage, it is worth to explicitly list the analogies and the  differences between  Eq.(\ref{FORMALDYN}) and the recent modeling framework introduced in \cite{KOENIG2016}. Both in Eq.(\ref{FORMALDYN}) of the present paper and in \cite{KOENIG2016}, the economic growth dynamics is measured via the log-productivity of the firms, which are assumed to evolve according to discrete time Markov chains. The innovation mechanism is characterized by a constant jump rate in both these modeling frameworks\footnote{In Eq.(\ref{FORMALDYN}), innovation is summarized by the parameter $\epsilon$, while it is denoted by $p$ in \cite{KOENIG2016} (see Section 4.1).}. The basic differences between Eq.(\ref{FORMALDYN}) and the approach derived in \cite{KOENIG2016} are the following:
\begin{itemize}
\item[\textit{(1)}] In \cite{KOENIG2016}, the imitation and innovation mechanisms  do not simultaneously coexist. Firms either imitate or innovate, and the alternations between attitudes are endogenously triggered by the value of an underlying utility function. In Eq.(\ref{FORMALDYN}) however, we assume that both the innovation and imitation processes steadily co-exist. Let us observe that in \cite{BENHABIB2017}, innovation and imitation (which is referred as adoption) mechanisms do also coexist. 

\vspace{0.2cm}
  \item[\textit{(2)}] In \cite{KOENIG2016}, the Markov chain  is not restricted to be a BD process. Instead, imitation is assumed to enhance the log-productivity via  jumps of random lengths. For a given firm ${\cal A}_k$, the length of the jumps depends on the log-productivity distance between  ${\cal A}_k$ and a randomly chosen leader ${\cal A}_j$ for $j\neq k$. In Eq.(\ref{FORMALDYN}), we assume that the log-productivity  follows a BD process where the jumps are limited to a single echelon by time step $\Delta$. Any firm ${\cal A}_k$ observes in real-time and within an observation range $(U \, a)$, the log-productivity states of all her concurrents, and it is the number of found leaders ${\cal N}_k(t)$ which triggers the  jump rate towards the log-productivity improvements. 

\vspace{0.2cm}
  \item[\textit{(3)}] In a single  time step $\Delta$, the imitation mechanism in Eq.(\ref{FORMALDYN}) does not lead to a log-productivity increase with certainty, rather imitation incidences are stylized by a jump probability bias. In the present approach, we have that $ p(ka, \nu \Delta,  \vec{X}(\nu\Delta))+ q(ka, \nu \Delta,  \vec{X}(\nu\Delta)) \equiv 1$. This extra probability-conservation law will ultimately enable us to derive exact results.
 \end{itemize}
\noindent  Let us now us reorganize the terms of Eq.(\ref{FORMALDYN}) in order to reach a more suggestive form\footnote{Let us observe that the nonlinear BD dynamics given in Eq.(\ref{FORMALDYN}) is basically a generalization of the so-called {\it clannish random walk} (see Chapter 2, Section 7 in \cite{MONTROLL1979}).}:
\begin{equation}
\label{FORMALDYN2}
\begin{array}{l}
P\left[(ka, (\nu+1)\Delta \right] - P\left[(ka, \nu \Delta \right] = {1 \over 2} \left\{P\left[(k+1)a, \nu \Delta ) \right]  -2 P(ka, \nu \Delta)  + P\left[(k-1)a, \nu \Delta ) \right]  \right\} +  \\ \\

\qquad \qquad { \alpha \over 2}  \left\{P\left[\left(k-1)a, \nu \Delta \right) \right]   -   P\left[(k+1)a, \nu \Delta ) \right]\right\} +  \\ \\ 

{\beta \over 2} \left\{ \left(\sum_{m= k-1}^{m = k-1+U} \left[ P(ma, \nu \Delta)\right] P\left[\left(k-1)a, \nu \Delta \right) \right]\right) -\left(  \sum_{m= k+1}^{m = k+1+U} \left[ P(ma, \nu \Delta)\right] P\left[\left(k+1)a, \nu \Delta \right) \right] \right) \right\}
\end{array}
\end{equation}
\noindent We perform now the continuous state-space-and-time limit, $\Delta \rightarrow 0$ and $a \rightarrow 0$, for which we can write $P(ka, \nu \Delta)  \mapsto \rho(x,t)$ with $x \in \mathbb{R}$ and $t \in \mathbb{R}^{+}$. When  $\Delta  \rightarrow 0$ and $a\rightarrow 0$, we proceed via the standard limiting procedure\footnote{See Footnote $4$, and also  Section 2.A in \cite{DIXIT1994}.} by simultaneously imposing that:
\begin{equation}
\label{LOMOSS}
\left\{
\begin{array}{l}

\alpha \sim \sqrt{\Delta}, \\  \\

a\sim\sqrt{\Delta}, \\ \\ 
\beta \sim \sqrt{\Delta}.
\end{array}
\right.
\end{equation}
\noindent  Using the identities:
\begin{equation}
\left\{
\begin{array}{l}
g(x+a) -g(x-a) = \sum_{k=0}^{\infty}  {a^{2k+1}\over k! } {d^{(2k+1)} \over dx^{2k+1} } \left[g(x)\right], \\ \\
g(x+a) + g(x-a) = \sum_{k=0}^{\infty}  {a^{2k}\over k! } {d^{(2k)} \over dx^{2k} } \left[g(x)\right],
\end{array}\right.
\end{equation}
\noindent the  Taylor expansion of  Eq.(\ref{FORMALDYN2}), up to the first order in $\Delta$, and to the second order in $a$, enables us to rewrite:
\begin{equation}
\label{CLIMIT}
\partial_t \rho(x,t) = \underbrace{\left({a^{2} \over 4 \Delta} \right)\partial^{2}_{xx}  \rho(x,t) - \left( {a \alpha \over 2 \Delta} \right) \partial_x\rho(x,t)}_{d(x,t)}-
\underbrace{\left({ \beta a\over \Delta} \right) \partial_x \left\{\left[\int_{x}^{x+U}  \rho(y,t) dy  \right] \rho(x,t) \right\}}_{\Psi_u(x,t)}.
\end{equation}
\noindent Note that, besides being differentiable once with respect to time $t$ and twice with respect to the variable $x$, the probability interpretation of $\rho(x,t)$  imposes that  $\int_{{\mathbb R}} \rho(x,t) dx =1$. In Eq.(\ref{CLIMIT}), one recognizes  a purely diffusive part $d(x,t)$, the origin of which can be directly traced back from the innovation and a nonlinear and nonlocal component  $\Psi(x,t)$ describing the imitation mechanism. The nonlinear and nonlocal Fokker-Planck equation  Eq.(\ref{CLIMIT}) is the basic deterministic\footnote{In the mean-field approach, thanks to the law of large numbers, explicit randomness disappears from the description. It is  effectively taken into account by the diffusive part of the dynamics.} evolution to be studied in the present paper. 

\vspace{0.5cm}
\noindent{\it Remarks}. 
\begin{itemize}
\item[\textit{(1)}] When pure innovation is considered, namely  $\alpha=0$ in  Eq.(\ref{CLIMIT}),  we exactly recover the dynamics studied in \cite{KOENIG2016}\footnote{In \cite{KOENIG2016}, the innovation parameter is denoted by $p$ and in the limit of a large population, the innovation process is purely diffusive and reads as $d(x,t)$.}.

\vspace{0.2cm}
 \item[\textit{(2)}] In absence of innovation, namely when $d(x,t)=0$ and for the infinite observation range capability ( $U \, a\rightarrow \infty$), we observe that Eq.(\ref{CLIMIT}) reduces to :
\begin{equation}
\label{CLIMIT1}
\left\{
\begin{array}{l}
\partial_t \hat{G}(x,t) = \left({\beta a\over \Delta} \right) \partial_x \left\{ - \hat{G}(x,t) +{1 \over 2}\hat{G}^{2}(x,t) \right\}, \\ \\
1-\hat{G}(x,t) := \int_{x}^{\infty}\rho(y,t) dy. 
\end{array}
\right.
\end{equation}
\noindent Up to a rescaling, we note that Eq.(\ref{CLIMIT}) reproduces the $q=1$ long-range limit of the modeling framework exposed in \cite{KOENIG2016}. Accordingly, it  is also identical to the knowledge growth dynamics model pioneered by Lucas in \cite{LUCAS2009}.
\end{itemize}

\vspace{0.7cm}
\noindent When $U>0$ and $\gamma > 0$, we emphasize that in Eq.(\ref{CLIMIT}), the density $\rho(x,t)$  obeys a class of  nonlinear and non-local partial differential equations  for which exact solutions are not to be expected in full generality. However, for limiting regimes, explicit solutions can be analytically worked out.

\vspace{0.2cm}
\begin{itemize}
  \item[{\it (A)}]  {\bf Infinitesimal Imitation Range} (Cautious Agents)\\

\vspace{-0.4cm}
 \noindent  This regime assumes that interactions are strictly limited to an infinitesimal spatial range $U$. This allows us to Taylor-expand (up to first order in $U$)  the integral term  in Eq.(\ref{CLIMIT}) to obtain:
  
  \begin{equation}
\label{TAYGENBUR2}
\left\{
\begin{array}{l}
\partial_{t} \left[ \rho(x,t) \right]= -\partial_x \left\{\left[ \alpha + (\gamma U) \rho(x,t) \right] \rho(x,t) \right\} + {\sigma^{2} \over 2 }\partial^{2}_{xx} \left[ \rho(x,t)\right],\\ \, \\
\lim_{|x|\rightarrow \infty} \left[\rho(x,t) \right] =0.
\end{array}
\right.
\end{equation}

\vspace{0.4cm}
\item[{\it (B)}] {\bf Infinite Imitation Range} (Enterprising Agents)\\
  
\vspace{-0.4cm}
For the extreme opposite case to regime {\it (A)}, we can again explicitly work out the dynamics in a differential form. Instead of the density $\rho(x,t)$, which is involved in Eq.(\ref{TAYGENBUR2}), let us  introduce and focus here on the complementary distribution function $G(x,t)$ with strictly negative partial derivative with respect to $x$:
  \begin{equation}
\label{NOTE}
G(x,t) = \int_{x} ^{\infty} \rho(y,t) dy \qquad \Rightarrow \qquad \partial_{x} G(x,t) = - \rho(x,t).
\end{equation}

\noindent When $U= \infty$, using the notation of  Eq.(\ref{NOTE}) allows us to rewrite  Eq.(\ref{CLIMIT}) as: 
\begin{equation}
\left\{
\begin{array}{l}
\label{EXPGENBUR}
\partial^{2}_{x,t} \left[G(x,t)\right]  = -\partial_{x} \left\{\left[  \alpha + \gamma G(x,t) \right] \left(\partial_{x} G(X,t)\right) \right\} + {\sigma^{2} \over 2} \partial^{3} _{xxx}\left[ G(x,t)\right],\\ \,\\
G(-\infty,t) =1 \qquad {\rm and} \qquad G(+\infty,t) =0.
\end{array}
\right.
\end{equation}
\noindent By integrating Eq.(\ref{EXPGENBUR}) once with respect to $x$  and imposing a  vanishing integration constant (we effectively assume as usual  that no probability current flows at infinity),  we immediately obtain:
\begin{equation}
\left\{
\begin{array}{l}
\label{EXPGENBUR1}
\partial_{t} \left[G(x,t)\right]  = - \left\{\left[  \alpha+ \gamma G(x,t) \right] \partial_{x} G(X,t) \right\} + {\sigma^{2} \over 2} \partial^{2} _{xx} \left[G(x,t)\right],\\ \, \\
G(-\infty,t) =1 \qquad {\rm and} \qquad G(+\infty,t) =0.
\end{array}
\right.
\end{equation}
\end{itemize}

\vspace{0.2cm}
\noindent Except for their boundary conditions, we observe that  Eqs.(\ref{TAYGENBUR2}) and (\ref{EXPGENBUR1}) coincide. To  solve these partial differential equations (PDEs), first we introduce the change of referential: $x\mapsto z = \left[ x-\alpha t\right]$ and accordingly, Eqs.(\ref{TAYGENBUR2}) and (\ref{EXPGENBUR1})  can be rewritten as:
\begin{equation}
\label{BURGO}
\partial_{t} \left[ \varphi(z,t) \right]= - \Gamma \partial_z \left[ \varphi(z,t) \right] ^{2}  + {\sigma^{2} \over 2 }\partial^{2}_{zz} \left[ \varphi(z,t)\right],
\end{equation}

\noindent where the parameter $\Gamma$  in Eq.(\ref{BURGO}) is suitably identified as:

\begin{equation}
\label{GAMMA}
\Gamma = 
\left\{
\begin{array}{l}
\frac{\gamma U}{2} \,\, {\rm and}\,\, \varphi (z,t) := \rho(z,t) \quad\,\, {\rm for \,\, the \,\, model  \,\, given \,\, in \,\, Eq.(\ref{TAYGENBUR2}) },\\\,\\
\,\, \frac{\gamma}{2}  \,\,\, \,\, {\rm and}\,\, \varphi (z,t) := G(z,t)  \quad  {\rm for\,\, the \,\, model  \,\, given \,\, in \,\, Eq.(\ref{EXPGENBUR1}). }
\end{array} 
\right.
\end{equation}

\noindent Hence, for $\varphi(z,t)$, we see that Eq.(\ref{BURGO}) is the  Burgers' equation, which can be linearized by  using a logarithmic  transformation, and hence explicit solutions are well-known. Dependent on their boundary conditions, these explicit solutions of Eq.(\ref{BURGO}) are recalled in {\it (A)} and {\it (B)} below.

\vspace{0.3cm}
\begin{itemize}
  \item[{\it (A)}]  {\bf Infinitesimal Imitation Range} (Cautious Agents),  {\it c.f.} Eq.(\ref{TAYGENBUR2}) \\
  
\vspace{-0.45cm}
For the boundary condition $\lim_{|x|\rightarrow \infty} \left[\varphi(z,t) \right] =0$ and for the initial condition $\varphi(z,0) = F(z)$, the solution of Eq.\eqref{BURGO}, and subsequently the agent density  function solving the model described by Eq.(\ref{TAYGENBUR2}), is given by (see Eq.(8.4.14) in \cite{DEBNATH2005}):
  \begin{equation} \label{GENOBURO}
\rho(z,t)  = \varphi(z,t) = {\int_{\mathbb{R}} \left({z -\zeta \over 2 \Gamma t} \right) e^{- \left( {f \over 2 \nu }\right) } d\zeta \over \int_{\mathbb{R}} e^{- \left( {f \over 2 \nu }\right)}  d\zeta},
\end{equation}

\noindent with the definitions:
\begin{equation}
\label{DEFOSS}
\nu = {\sigma^{2} \over 4 \Gamma} \qquad {\rm and} \qquad f= f( \zeta, z,t) = \int_0^{\zeta} F( y) dy + {(z -\zeta)^{2} \over 2 \Gamma t}.
\end{equation}

\noindent In particular, in the presence of small noise intensity and for the initial condition $\varphi(z,0) = F(z)= \delta(z) \Theta(z)$, the  asymptotic behavior ({\it i.e.} $t \rightarrow \infty$) of the dynamics given by Eq.(\ref{GENOBURO}) can be approximately written as:

\begin{equation} \label{GENOBUROSS}
\rho(z,t)  =\varphi(z,t) \simeq \left \{ 
\begin{array} {l} 
{z \over 2 \Gamma t} = \frac{z}{\gamma U t}  \quad {\rm if\, \,}0<z< \sqrt{4 \Gamma  t},\\ \,\\ 0 \quad {\rm \,\,\,\,\,otherwise},
\end{array} \right.
\end{equation}

\noindent which, for this vanishing noise regime,  converges toward a shock wave-like pattern, as shown in Figure \ref{graph_burger}. 
\begin{figure}[h]
\begin{center}
\hspace{-0.2cm}
\includegraphics[width=169mm, height=65mm]{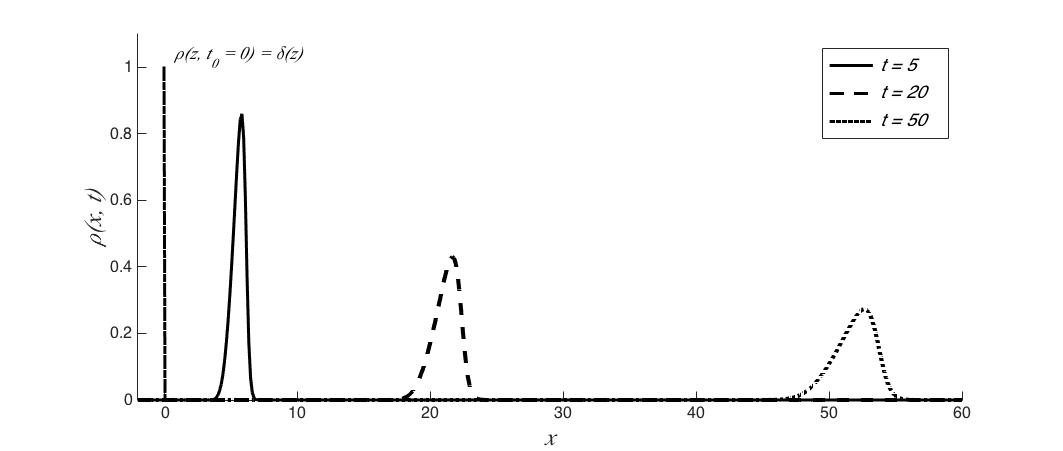}
\caption{Collective dynamics observed for the infinitesimal imitation range as given by Eq.(\ref{GENOBURO}), when $\gamma=1$,  $\alpha=1$, $\sigma=0.2$, and $U=0.1$. The interactions between the agents produce an asymmetric shape for the density $\rho(x,t)$, which propagates at speed $(\alpha + \gamma U/2)$. Diffusion precludes the formation of a stationary dynamic  pattern. Thus, for $t \rightarrow \infty$, the density $\rho(x,t)$ flattens while remaining normalized to unity. Ultimately, the productivity states tend to be widely dispersed ({\it i.e.} absence of flocking). The short-range imitation mechanism precludes the productivity leaders to give rise to a stable growing productivity wave.}
\label{graph_burger}
\end{center}
\end{figure}

\vspace{0.2cm}
\item[{\it (B)}] {\bf Infinite Imitation Range} (Enterprising Agents), {\it c.f.} Eq.(\ref{EXPGENBUR1})\\
    
\vspace{-0.4cm}
In this case, the boundary conditions are equal to $\varphi(-\infty,t) =1\,\,  {\rm and} \,\, \varphi(+\infty,t) =0$. For any arbitrary initial condition $\varphi(z,0) = F(z)$, the solution of Eq.\eqref{BURGO}, and subsequently the agent probability distribution that solves the model described by Eq.(\ref{EXPGENBUR1}), can be written for asymptotic time as the following traveling wave solution (see Eq.(8.3.8 )in \cite{DEBNATH2005}):
\begin{equation}\label{BUROTANH}
\varphi(z,t) = {1 \over 2}\left[ 1 -\tanh \left({\Gamma(z - \Gamma\, t )\over \sigma^{2}}\right)\right], \qquad \tanh (x)=\frac{e^x-e^{-x}}{e^x+e^{-x}}=\frac{\sinh (x)}{\cosh (x)}.
\end{equation} 
\noindent 
Using Eq.\eqref{NOTE}, by differentiating Eq.\eqref{BUROTANH}, we determine that the agent density function solving the model described by Eq.(\ref{EXPGENBUR1}) is a soliton-like propagating wave, as shown in Figure \ref{travelling_wave}:
\begin{equation}
\label{ADENS}
\rho(z,t) = - \partial_z \varphi(z,t) = {\Gamma \over 2 \sigma^{2} \cosh^{2} \left({\Gamma(z - \Gamma\, t )\over \sigma^{2}}\right)} =  {\gamma \over 4 \sigma^{2} \cosh^{2} \left({\gamma(2z - \gamma\, t )\over 4\sigma^{2}}\right)}.
\end{equation}

Observe that the larger the noise amplitude $\sigma$, the flatter the resulting solution.

\end{itemize}

\begin{figure}[h]
\begin{center}
\hspace{-0.2cm}
\includegraphics[width=169mm, height=65mm]{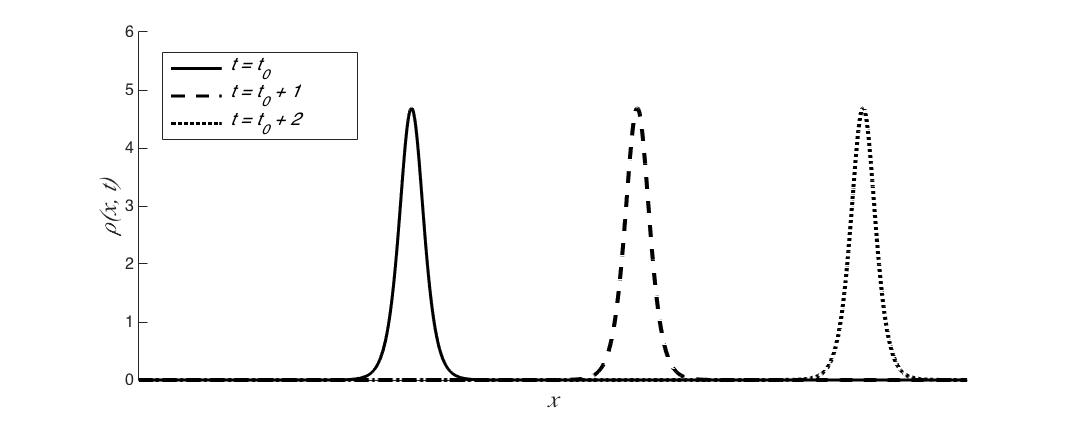}
\caption{Collective dynamics observed for the infinite imitation range, as given by Eq.(\ref{ADENS}), when $\gamma=1$, $\alpha=1$, and $\sigma=0.2$. The imitation mechanism generates a collective productivity wave with constant variance,  which travels at constant velocity $(\alpha +\gamma/2)$, without shape alteration (the transient evolution is not represented in this figure). Thus, the agents remain spatially tuned together ({\it i.e.} presence of flocking). They collectively progress on the abstract  productivity real line with constant dispersion. The large-range imitation mechanism favors the influence of the  leaders and  ultimately generates a  collective spatio-temporal pattern.}
\label{travelling_wave}
\end{center}
\end{figure}

\noindent {\it Remark}: 
\begin{itemize}
\item[] As shown by Eq.(\ref{ADENS}), and contrary to the reaction-diffusion evolutions such as the equation derived in \cite{STALEY2011}, the collective log-productivity growth rate $\gamma$ does not depend on the amplitude of the diffusion $\sigma$, which affects only the shape of the traveling wave.\footnote{Due to the so-called Rankine-Hugoniot relation, it is known that for scalar hyperbolic conservation laws, to which the Burgers Eq.(\ref{BURGO}) belongs, the propagating speed of the traveling wave depends only on the boundary values of $\varphi(-\infty,t) =1$  and $\varphi(+\infty,t) =0$ (see \cite{BEDJAOUI2004}).}.
\end{itemize}

\vspace{0.3cm}
\noindent  It is interesting to observe the fundamentally different  dynamic  behaviors emanating in the two regimes {\it (A)} and {\it (B)} exposed above, the solutions of which are given  by Eqs.(\ref{GENOBURO}) and (\ref{ADENS}). The variances $s^{2}(t)$ associated with Eqs.(\ref{GENOBURO}) and (\ref{ADENS}) can be explicitly written as:
\begin{equation}
 \label{VAROOS}
 s^{2} (t)= \int_{\mathbb{R}} \left(z^{2} \right)\rho(z,t)\,  dz=\left\{
\begin{array}{l} 
{1 \over 6} (\Gamma t)\qquad \,\,\,{\rm for \,\, short\,\,  imitation \,\,range,\,\,}c.f. \,\, {\rm case\, }(A), \\ \, \\
{\pi^{2} \sigma^{4} \over 3 \gamma^{2}} {\sigma^{4} \over \Gamma^{2}} \quad\,\,\,\,\, {\rm for \,\, large\,\,  imitation \,\,range,\,\,}c.f. \,\, {\rm case\, }(B),
\end{array}
\right.
\end{equation}
\noindent Eq.(\ref{VAROOS})  exhibits  a structural change for agents that behave with  short versus long imitation ranges. Only long-range  imitation mechanisms  sustain the emergence of stable stationary traveling waves (soliton-like) with constant variances. From shorter range mimicry, the emergent dynamic pattern is dominated by   diffusion. In that case,  the variance of the productivity wave grows with time and  ultimately leads to an evanescent dynamic  pattern ({\it i.e.} no stable constant variance productivity wave can survive). These two drastically different productivity evolutions suggest that there should exist a critical imitation strength below which the stable dynamic growth pattern cannot survive. This issue is addressed in the next section.\\
\\
The growth rates that emerge in these regimes are equal to \textit{(A)} $\alpha + \frac{\gamma U}{2}$ ($U$ small) and \textit{(B)} $\alpha + \frac{\gamma}{2}$. In both cases, the engine of growth is composed of 2 components: The first term ($\alpha$) is the result of individual attempts toward innovation, and the second term ($\frac{\gamma U}{2}$ or $\frac{\gamma}{2}$) is the consequence of mutual interactions. Aligned with the findings exposed in \cite{LUTTMER2012a}, this suggests that innovation and imitation together are ultimately required to create a balanced growth path. 


\subsection{Crowd-Based Agents' Interactions}\label{sct_mod}

\noindent We have thus far focused on limiting cases that involve infinitesimally short and infinitely long-range imitation mechanisms. We now broaden this framework by considering the following generalization. For an infinite observation range, we extend our mutual interaction rule by introducing a symmetric weighting ${\cal G}(x - \langle X(t) \rangle ) = {\cal G}( \langle X(t) \rangle -x)$ factor that depends on the remoteness of each agent with the swarm barycenter $\langle X(t) \rangle $. When ${\cal G}$ is a decreasing function of its argument, it will generate a {\it conformist tendency} as the agents attach more importance to average behavior. Conversely, for increasing ${\cal G}$'s, agents are more influenced by leaders or laggards. As in  our model, the imitation mechanism  systematically  depletes the laggard population in favor of the leader population,  and increasing the ${\cal G}$-modulation effectively describes the strong influence of the frontier technology leaders. To summarize, we assume as before that agents systematically tend to imitate their leaders, but we modulate the strength of imitation with the idea that leaders who are far away are either more or less  influential than those who are close to the crowd barycenter.  Assuming once again the validity of the mean-field  approach and choosing $\gamma=1$, we now generalize the interaction kernel given in Eq.(\ref{CLIMIT}) by writing:

{\footnotesize
\begin{equation}
\label{GENBURMO}
\left\{ 
\begin{array}{l}
\partial_{t} \left[ \rho(x,t) \right]= -\partial_x \left\{\left[ \alpha + \underbrace{\int_{x}^{\infty} {\cal G} \left[y- \langle X(t) \rangle \right] \rho(y,t) dy}_{{\rm  imitation \,\, modulated \,\, drift}}\right] \rho(x,t) \right\} + {\sigma^{2} \over 2 }\partial^{2}_{xx} \left[ \rho(x,t)\right], \\ 
\,\\
\rho(x,t) \in [0,1]\qquad {\rm and} \qquad \lim_{|x|\rightarrow \infty} \left[\rho(x,t) \right] =0,
\end{array}
\right.
\end{equation}
}

\noindent where $\langle (X(t) \rangle :=  \int_{\mathbb{R}} x \rho(x,t) dx$ is the swarm barycenter. Let us  emphasize that in Eq.\eqref{GENBURMO}, the imitation range is effectively infinite (\textit{i.e.} the integral boundary is $+ \infty$). As in Section 2.1, we would like to investigate the possible existence of a stationary density with constant variance and traveling velocity, namely, a solution of the form $\rho(x-vt)$. In general, the nonlinear
and nonlocal character of Eq.(21) precludes us from finding an explicit analytical solution.
However, as shown in Appendix C, the specific choice\footnote{When $\eta=0$ (and the observation range is infinite), the dynamics corresponds to the one studied in \cite{IWAI1984}. Indeed, the imitation process matches in this case the situation where each agent's imitation activity consists of randomly observing one of her peers, and replicating the observed productivity (by augmenting her drift) as long as it is higher than her own productivity.}:

\begin{equation}
\label{CHOICENICE}
\left\{
\begin{array}{l}
{\cal G}(x) = {\cal A}(\eta, \sigma^{2}) \cosh^{-\eta}(x), \\ \\
{\cal A}(\eta, \sigma^{2}) =  {(2- \eta) [\Gamma(1 - {\eta \over 2}) ]^{2}\over  2^{\eta} \Gamma(2- \eta)}  \, \sigma^{2}  , \qquad \eta\in [-\infty,2[,
\end{array}
\right.
\end{equation}

\noindent leads to the explicit constant variance productivity wave growth:

\begin{equation}
\label{PROLONG}
\left\{
\begin{array}{l}
\rho(x ,t)= {\cal N}[(2-\eta)] \cosh^{(\eta-2)}(x- (\underbrace{\alpha + w}_{v})t),  \qquad \eta\in [-\infty,2[\\ 
\\
w=  (2 - \eta) {\sigma^{2} \over 2}
\end{array}
\right.
\end{equation}

\noindent where ${\mathcal N}[(2-\eta)]$ is the  normalization factor of the density $\rho(x,t)$. The traveling velocity includes two components, namely $\alpha$ which is due to the innovation rate and $w$ that is due to the mutual interactions. Observe that the smaller is the $\eta<0$, the more peaked the emerging soliton becomes. Conversely, for $0<\eta \lesssim 2$,  a {\it  table-top} soliton is obtained.\\ 

\noindent From Eq.(\ref{PROLONG}), three different regimes can be distinguished depending on the value of $\eta$:

\vspace{0.1cm}
\begin{itemize}
\item[\textit{(1)}] \textbf{Cautious Agents.}\\
When the control parameter $\eta \in [2, \infty[$, the effective interaction strength is too limited to give rise to flocked collective behavior. The decay exhibited by the function $\mathcal{G}$ is strong, implying that only leaders close to the swarm barycenter affect the dynamics. This stylizes cautious behaviors where strong conformism dominates and where productivity leaders have a negligible influence on their peers. Accordingly, in this situation, no flocking traveling soliton wave can be sustained, and only a diffusive time-evanescent wave results (with growing variance).
\end{itemize}
\noindent Conversely for $\eta \in ]- \infty , 2]$, flocking soliton waves emerge, and two distinctive strategies can be highlighted.
\begin{itemize}
\item[\textit{(2.a)}] \textbf{Weakly Enterprising Agents.}\\
For $\eta =[0,2[$, a cautious attitude still dominates, as the function $\mathcal{G}$ given by Eq.(\ref{CHOICENICE}) exhibits a slow decay (remember that $\cosh^{-\eta}(x) \simeq (1/2)e^{ -\eta x}$  for $x \rightarrow \infty$), meaning that remote leaders, while still influencing the dynamics, are given an importance that decreases remotely. This stylizes a relatively moderate enterprising behavior as agents are ready to take into account outliers, but with reduced influential power.  
\item[\textit{(2.b)}] \textbf{Strongly Enterprising Agents.}\\
For $\eta < 0$, the pace is given by leaders located close to the productivity frontier (this results from the asymmetry of the model), which highlights a net progressive-oriented attitude. Agents pay more attention to the productivity leaders than to their fellow agents situated close to the crowd barycenter. This produces a decrease in the tail of the agents' distribution and thus sharpens the soliton wave.
\end{itemize}

\vspace{0.3cm}
\noindent {\it Remark}: 
\begin{itemize}
\item[] The evolution described by Eq.(\ref{PROLONG}) has to be contrasted with the result derived in Section 5.2 of \cite{BENHABIB2017} (\textit{i.e.} absence of \textit{excludability}) where a so-called distortion coefficient called $\kappa$ turns out to play a similar role than $\eta$ in Eq.(\ref{CHOICENICE}). It is observed in \cite{BENHABIB2017} that $\kappa$ influences the shape of the traveling wave but not its speed, whereas in Eq.(\ref{PROLONG}) the speed of the productivity growth also depends on $\eta$. While in both approaches the parameter $\eta < 0$ (respectively $\kappa$) confers a strong driving influence to the leaders close to the productivity frontier, there exists a drastic difference between the frontier dynamics itself. In our modeling framework, the innovation frontier is described by a diffusion process, therefore with a transition probability density obeying a parabolic Fokker-Planck evolution, with PDE characteristics propagating at infinite speed (\textit{n.b.} this is due to the underlying existence of unbounded diffusive excursions). The unboundedness of the characteristics speed ultimately affects the global productivity growth velocity. This has to be contrasted with the dynamics studied in \cite{BENHABIB2017}, where the frontier evolution is driven by a two-state, continuous-time, Markov chain, hence leading to transition probability densities obeying hyperbolic Fokker-Planck evolutions with strictly finite velocity characteristics. Due to these characteristics, the driving influence of the frontier is weaker and the $\kappa$-distortion is observed to affect the shape but not the speed of the log-productivity growth wave.
\end{itemize}

\vspace{0.4cm}
\noindent \textbf{Phase Transition}\\
\noindent Therefore, for the modulation choice given by Eq.(\ref{CHOICENICE}), the critical decay threshold $\eta =2$ is a bifurcation  parameter, which separates two drastically different growth regimes. When $\eta>2$, growth cannot be sustained as the mutual interactions are too limited. Conversely, when $\eta<2$, the imitation strength is large enough to trigger a stationary balanced growth path with strictly positive growth. This aspect has not been unveiled yet in the literature. \\

\begin{figure}[h]
\begin{center}
\hspace{-0.2cm}
\includegraphics[width=165mm]{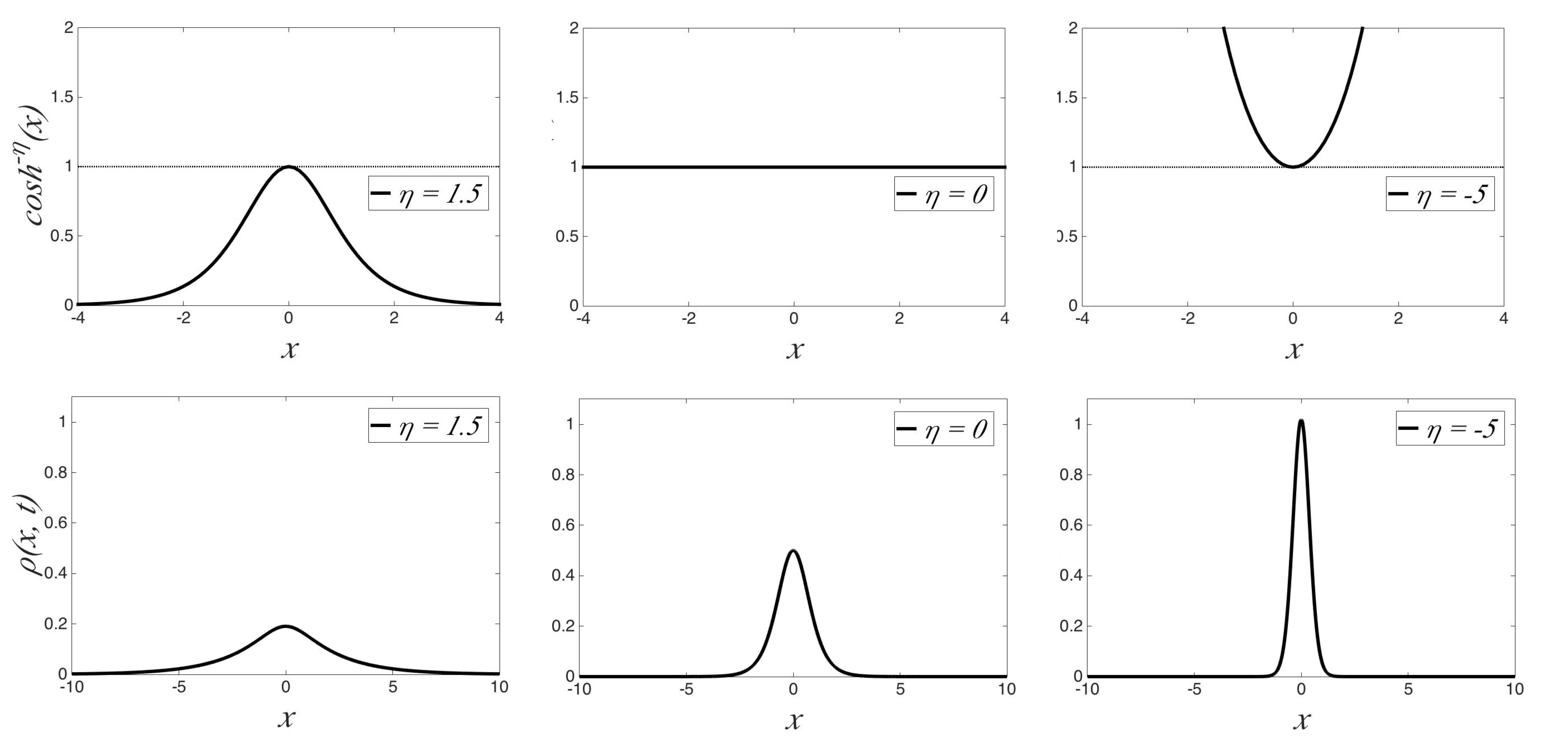}
\caption{Barycentric modulation functions $\cosh^{-\eta}(x)$, for different values of $\eta$, and corresponding collective productivity wave $\rho{x,t}$.}
\label{soliton_comparison}
\end{center}
\end{figure}

\noindent \textbf{Agent Dispersion and Inequalities}\\
As shown in Figure \ref{soliton_comparison}, 
the dispersion of the agents in the stationary state decreases with the strength of interaction in the economy (\textit{i.e.} when $\eta$ decreases). A higher degree of imitation, jointly with larger influence of the productivity leaders, decreases the inequality level between the agents\footnote{Note that the stationary agents' distribution exhibits an exponential tail. This behavior has to be contrasted with the one exposed in \cite{LUTTMER2012a}, where a fat tail emerges.}.\\

\vspace{-0.3cm}
\subsection{Relation with the Schumpeterian Literature}

The objective of this section is to connect our modelling framework with former related classical economic contributions devoted to economic growth.

\subsubsection{Extra Drift Parameter $\boldsymbol{\alpha}$ and Solow Residual}

\noindent In our basic microscopic modelling defined in Eq.\eqref{FORMALDYN}, we introduce a  bias $\alpha$  in the Markov chain transitions. This gives rise to an extra drift in the subsequent continuous dynamics, namely Eqs.\eqref{CLIMIT}, \eqref{TAYGENBUR2}, \eqref{EXPGENBUR1}, and \eqref{GENBURMO}. Let us now connect the parameter $\alpha$ with the macro-economic variables as exposed in \cite{AGHION2009}. From a  Cobb-Douglas production framework, \cite{AGHION2009} shows how the aggregate instantaneous growth rate of output per person $g(t)$ can be attributed to two separate contributions. The first one is technological progress, namely the {\it total factor productivity } (TFP) $A(t)$. The second one is factor accumulation. Following \cite{AGHION2009}, we have:
\begin{equation}
\label{COBB}
\left\{
\begin{array}{l}
Y(t) = A(t)   K^{a} (t) L^{1-a}(t) \quad \Rightarrow \quad  \underbrace{ \log\left[  {Y(t) \over L }\right] }_{:=\int_{t_0}^{t}gs)ds} = \underbrace{\log[A(t)]}_{.= X(t) }+  a\underbrace{ \log \left[{ K(t) \over L(t)}  \right]}_{:=\log(k(t)}  \\ \\ 
dX_t=   g (t) dt   -
a \left[  d\log(k(t)\right] dt, 
\end{array}
\right.
\end{equation}

\noindent where  $a\in [0,1]$ and $a  \left[  d\log(k(t)\right] $ stands for the \textit{capital-deepening} contribution. The so-called Solow residual $X_t$ is derived from a Cobb-Douglas evolution and hence it is a macro-economic variable with a purely deterministic evolution\footnote{In a macroscopic description, random fluctuations around  average paths are omitted. The law of large numbers implies that the aggregation of numerous microscopic evolution into a single macro-variable wipes out the fluctuating  contributions.}. Hence, the evolution described by Eq.(\ref{COBB}) has to be connected with the average $g(t)  = \int_{\mathbb{R}} x \rho(x,t) dx$ and with the probability density of the logarithm of the TFP $\rho(x,t)$. Focusing on stationary regimes, we  have $g(t) \mapsto g = (\alpha + \omega)$, where $\alpha$ isolates all Solow residual contributions that are not imputable to the imitation mechanism.

\subsubsection{Choice of the Noise Sources}

\noindent In our nominal  microscopic modelling given by Eq.\eqref{MFAPP}, the set of diffusion processes is driven by independent White Gaussian noise sources. Similarly, the underlying Markovian dynamics on  the  set of productivity ladders, as given by Eq.\eqref{FORMALDYN}, is characterized by independent jumps performed by the agents. In our dynamics, randomness is introduced to stylize Hicks' neutral productivity changes\footnote{A change is considered to be Hicks neutral if it does not impact the balance of labor and capital in a production function.}. Generally, sources of randomness can be multiple and, in particular, one may distinguish between an ubiquitous  {\it single exogenous} and {\it endogenous} noise sources:
\begin{itemize}
  \item[\textit{(i)}]  {\it  Single exogenous  stochastic process}, call it $\epsilon(t)$. It originates  from  the  environment  simultaneously shared by all agents. Specifically, random changes in the set of framework conditions of the economy will be perceived similarly by all firms.
  \item[\textit{(ii)}]  {\it  Endogenous stochastic processes}, call it $dW_{i}(t)$, for $i=1,2, \dots, N$. These processes are peculiar to each firm. Randomness due to management decisions, labor force characteristics such as skills, routines, learning and cognition (\cite{NELSON1982}), R\&D expenditures\footnote{Observe that fluctuations in the R\&D expenditure actually affect both endogenous and exogenous noise sources.}, probabilistic odds when ubiquitous failures occur in the production facilities or in the supply chains, all jointly affect the firms' TFP. The central limit theorem teaches us that the cumulative effects of such multiple random sources can be efficiently modeled by  colored\footnote{Colored noise processes have finite auto-correlations.} Gaussian stochastic processes. In the sequel, we will always assume that, for the time scales of interest, these Gaussian noise sources have vanishing auto-correlations and hence the endogenous $dW_{i}(t)$, $i=1,2, \dots, N$, are White Gaussian noise (WGN) processes. 
\end{itemize}

\noindent Accordingly, our dynamics structurally reads as:
\begin{equation}
\left\{ 
\begin{array}{l}
dX_{1}(t)  =\,\,\, [f(X_{1}(t),  \vec{X}(t))]dt + \epsilon(t) dt +  \sigma dW_{1}(t) , \\ 
dX_{2}(t)   =\,\,\, [f(X_{2}(t),  \vec{X}(t)) ]dt + \epsilon(t) dt +  \sigma dW_{2}(t), \\ 
\cdots \qquad \qquad \qquad \cdots \qquad \qquad \qquad \cdots  \\ 
dX_{N}(t)  =\,\,\, [f(X_{N}(t),   \vec{X}(t)) ]dt + \epsilon(t) dt+  \sigma dW_{N}(t).
\end{array} 
\right.
\end{equation}
\noindent For  $ i=1,2,\cdots, N$, the  $N$ aggregation processes  defined by $\xi_i(t) := \left[  \epsilon(t) dt +  \sigma dW_{i}(t)\right]$  are  clearly  cross-correlated\footnote{This is obviously due to the common component $\epsilon(t)$.}.  Since the $\epsilon(t)$ fluctuation process is simultaneously affecting  all components,  one may alternatively define: $\hat{f}(X_{k}(t),   \vec{X}(t) , \epsilon(t)) := [f( X_{k}(t) ,  \vec{X}(t) + \epsilon(t)]$. This last expression can be rewritten as:
\begin{equation}
\left\{ 
\begin{array}{l}
dX_{1}(t)  =\,\,\, [\hat{f}(X_{1}(t),  \vec{X}(t),\epsilon(t)) ]dt +  \sigma dW_{1}(t) , \\ 
dX_{2}(t)   =\,\,\, [\hat{f}(X_{2}(t),  \vec{X}(t), \epsilon(t)) ]dt +  \sigma dW_{2}(t), \\ 
\cdots \qquad \qquad \qquad \cdots \qquad \qquad \qquad \cdots  \\ 
dX_{N}(t)  =\,\,\, [\hat{f}(X_{N}(t)   \vec{X}(t), \epsilon(t))] dt+  \sigma dW_{N}(t) ,
\end{array} 
\right.
\end{equation}
\noindent where the $N$  endogenous noise sources $dW_{k}(t)$ are now independent as in our nominal dynamics given by Eq.\eqref{MFAPP}. Our basic goal being to study the collective interplay of innovation/imitation in the overall growth process, one would ideally like to isolate the contributions due to the agent interactions from those resulting from the commonly shared  exogenous environment. Namely, for a fixed realization of the environment process $\epsilon(t)$, one is interested in the resulting collective behavior of the whole swarm. In general, the intrinsic modelling nonlinearities preclude the possibility of such a clear separation analytically. We assume that $\epsilon(t)\simeq0$, as calibrated in our modelling, is valid either for small  exogenous noise sources or possibly  for slowly varying ones. In the latter case, $\epsilon(t)$ can indeed  be approximated by a piece-wise deterministic process with long average sojourn times in the successive random constant states $\epsilon_k$\footnote{When the average sojourn time is much larger than the  relaxation  time needed to reach equilibrium,  within one  period of the noise realization, we recover our nominal dynamics with $\alpha \mapsto \alpha + \epsilon(t)$.}. 

\subsubsection{Qualitative Effect due to Firm Entry and Exit}

\noindent Basically, our model assumes that entry and exit of firms are absent from the dynamics and so the population of firms $N$ is time-independent. Laggards always imitate leaders and never exit from the economy. This might look as a model weakness since entry and exit flows are definitely present in actual situations and hence naturally enter into many classical growth models, \cite{AGHION2009}.\\

\noindent Let us now assume that $N \mapsto N(t)$ to reflect the fact that, due to the presence of entry and  exit flows, the number of  firms is allowed to fluctuate with time. Indeed, technological laggards exit from the economy and firms close to the technological frontier are allowed to join. Accordingly, the corresponding modelling framework would require incorporating moving probability sinks (respectively probability sources), located near the laggards (respectively in the leader neighbourhood). This additional complexity  destroys the conservative nature of the Fokker-Planck dynamics considered in this study\footnote{It is a continuity equation which preserves positivity and  the probability mass. These properties are essential to construct analytically solvable models, as those presented here.} and has not been addressed in our present approach. In the presence of such probability sinks and sources, let us assume the existence of a stationary regime characterized by a vanishing average imbalance between entries and exits, so that  $\mathbb{E} \left\{N(t)  \right\} = N$ remains constant. Compared to the nominal density $\rho(x,t)$ solving Eq.\eqref{CLIMIT}, such entry-and-exit flows are likely to induce an extra right-handed skewness in the total factor productivity probability density, let us call it  $\rho_{{\rm skew}} (x,t)$. The right-handed skewness weakens the probability weight carried by the left tail of  $\rho_{{\rm skew}} (x,t)$  compared to  the nominal left tail  of $\rho (x,t)$ and, since normalization is assumed to be preserved, this implies:
\begin{equation}
0 \leq \int_{x}^{\infty} \rho(y,t) dy  \leq  \int_{x}^{\infty} \rho_{{\rm skew}}(y,t) dy \leq 1.
\end{equation}
\noindent Hence, entry-and-exit would clearly impact the distribution of firms' productivity, as it would further increase the efficiency of an imitation mechanism of the type used in  Eq.\eqref{CLIMIT}, and hence further enhance the overall growth process.\\

\subsubsection{Fat Tail of the Total Factor Productivity Distribution}

\noindent Asymptotically, our soliton solution, as given by Eq.\eqref{PROLONG}, behaves as:
\begin{equation}
\label{FATTAIL}
\displaystyle \rho(x,t) = { {\cal N}  \over \cosh^{2-\eta} (x-vt) } \quad \simeq_{x \rightarrow \infty} \quad {\cal N}e^{(\eta-2)(x-vt)},
\end{equation}
\noindent for $\eta \in [- \infty,2[$ with $v= (\alpha + \omega)$, where $\omega$ is entirely due to the imitation process, and ${\cal N}$ is the normalization factor. At this stage, remember that we have defined   $X_t := \ln(A_t) $, where $A_t$ itself is the total factor productivity. Accordingly, to compare with the empirical results obtained in Figure 1 of \cite{KOENIG2016}, we have to calculate:
\begin{equation}
\label{KMOM1}
\begin{array}{l}
\mathbb{E} \left\{ \ln(A_t)  \right\}  =  {\cal N} (2 - \eta) \int_{\mathbb{R}^{+}}  { \left[ \ln(a) \right]\over \cosh^{2-\eta} (ln(A_t) -vt) } d\ln(A_t)  = vt,\\ \\
\sigma^{2}_{K}= \mathbb{E} \left\{ (\ln(A_t) -vt)^{2}\right\} =  {\cal N}(2- \eta) \int_{\mathbb{R}^{+}}  {\zeta^{2} \over \cosh^{2- \eta} (\zeta)} d\zeta) ={1 \over 4} \zeta (2,{2 - \eta \over 2}) \\ \\ 
 \zeta := (\ln(A_t)-vt) \qquad {\rm and} \qquad  {\mathcal N}^{-1} (2 - \eta)  =  {1 \over 2} 4^{\nu}{ [\Gamma(2- \eta)) ]^{2} \over \Gamma[2(2 - \eta)]}, 
\end{array}
\end{equation}
\noindent where $\zeta \left(2,{2 - \eta \over 2}\right)$ stands for the Hurwitz-zeta function (details for the underlying  calculations are provided in Appendix F). Focusing finally on  the probability  tail, we have:
\begin{equation}
\label{KFAT}
\begin{array}{l}
\rho(x-vt) dx \quad  \mapsto  \quad \rho(\ln(A_t)-vt) d[(\ln(A_t) ]\simeq{\cal N}(2 - \eta) e^{(\eta-2)(\ln (A_t)  - vt)} d [\ln(A_t)], 
\end{array}
\end{equation}
\noindent which in terms of $A_t$  exhibits a fat tail\footnote{As sketched in the Figure 1 of \cite{KOENIG2016}.} with  negative slope $(\eta -2)<0$.\\

\noindent Let us now directly refer to the results exposed in \cite{KOENIG2016} (see Appendix 2B in the supplementary material), we have:
\begin{itemize}
  \item[\textit{(a)}]  $\mathbb{E} \left\{ \ln(A_t)  \right\}  =  0.027$ 
\item[\textit{(b)}]  $\sigma_{K}  \in  [1.61, 1.67] \quad \Rightarrow \quad \sigma_{K}^{2} \in [2.5,2.78  ]$
\item[\textit{(c)}]  Slope of the right fat tail $\lambda_{R} = - 3.73$
\end{itemize}

\vspace{0.2cm}
\noindent In Appendix F, we show that  $\sigma_{K}^{2} \simeq (2-\eta)^{-1}$. Using entry \textit{(b)} above, we can determine the  corresponding factor $\eta$ as:
$$
\begin{array}{l}
\eta \in[- 1.6, -1.65]  \quad \Rightarrow \lambda_{R} = (2-\eta) \in [-3.6,-3.65], \\ \\
\end{array} 
$$
\noindent hence showing a good slope compatibility  of the right fat tail with the collected data in \cite{KOENIG2016}. Since  $\eta<0$, we are in the regime \textit{(2.b)} of Section \ref{sct_mod}, thus indicating a net tendency to belong to  the regime of enterprising agents. 
%

\vspace{0.3cm}
\noindent Finally, the soliton velocity, as given by Eqs.\eqref{PROLONG} and \eqref{COMPAT}, reads as:
$$
\alpha + {1 \over 2} ( 2- \eta) \sigma^{2} = 0.027.
$$
\noindent Accordingly, once the dynamics sets itself into the travelling soliton equilibrium, we observe that our model imposes a  {\it fluctuation-transport} relation connecting the drift $\alpha$ and the underlying endogenous noise source $\sigma$.

\subsubsection{Technical Adoption and the Parameter $\boldsymbol{\eta}$}

\noindent In our modelling framework, the  control parameter $\eta$ exogenously weights the relative influence of leaders in the imitation process, thus reflecting the relative capacity of firms to  absorb other firms' technologies. Specifically, $\eta$ incorporates:
\begin{itemize}
    \item[(1)] The underlying quality of the legal environment (i.e., patent protection system). In environments with well-protected intellectual property rights, the potentially high benefit offered by occupying a monopolistic position generates  strong  incentives to track the technological frontier, thus implying an ad-hoc choice of a negative $\eta$, or conversely. For example, countries where a strong patent law system is implemented will observe a lower tendency for imitation as compared to countries where this is not the case. Note that in our approach, and contrary to \cite{KOENIG2016}, we do not explicitly incorporate a utility function that would enable us to fix an optimal value for $\eta$.
    \item[(2)] The specific characteristics of the industry under consideration. As exposed in \cite{NELSON1982}, industries with different degrees of technological sophistication exhibit different behaviors with respect to imitation. For example, in science-based industries, the strongly protected intellectual property environment weakens the capability of imitation. The present paper hence draws on the building blocks provided by \cite{NELSON1982} and complements it with analytical tools.
\end{itemize}


\vspace{-0.5cm}
\subsubsection{\bf Auto-Catalytic Mechanism and Growth } 

\noindent We observe in  Eq.\eqref{TAYGENBUR2} that for this limiting infinitesimal interaction range, the imitation process effectively reduces to an auto-catalytic contribution given by  $\gamma U \rho^{2}(x,t)$. While such nonlinearity  is commonly encountered in chemistry \cite{NICOLIS1989}, it has been less remarked in the context of economics. Nevertheless, a similar mechanism has been clearly identified in \cite{SAVIOTTI1998}, where auto-catalytic nonlinearity is explicitly pointed out. Quoting  the authors: {\it [...] Imitation is an auto-catalytic phenomena in the sense that the higher the rate of imitation, the greater the incentive for other firms to imitate, at least to a certain point [...]}.


\vspace{0.3cm}
\subsection{Simulations and Model Generality}

To derive the exact results presented in Section 2,  
several analytical limitations have been imposed, namely: 
\begin{itemize}
\item[\textit{(a)}] the agent population is assumed to be very large (effectively  $N \rightarrow \infty$) for the mean-field approach to be strictly valid, 
\item[\textit{(b)}] the interaction range $U$ is either infinitesimally small or infinitely large,
\item[\textit{(c)}] the use of WGN to drive the evolution, 
\item[\textit{(d)}] the homogeneity of the agent population. 
\end{itemize}
This set of hypotheses is barely expected to be strictly realized in actual situations, and this raises naturally the question {\it regarding the robustness of our observations and conclusions under slight modifications of our basic hypotheses.} To discuss this fundamental issue, we report in Appendix E an extended set of simulations that illustrate that the analytical results in Section 2 are not qualitatively affected by (slightly) relaxing the hypotheses needed to derive the obtained exact solutions. 



\section{Endogenous Growth and Mean-Field Games}\label{sct_endo}

In the preceding section, we showed how the imitation mechanism influences the propagation of economic growth in a large population of interacting agents. While in Section \ref{sct_exo}, the agents' imitation strategy was exogenously defined, we now focus on MFGs, and more precisely on the collective dynamics that emerge  from 
an optimal control problem in which agents minimize an individual objective function. In the MFG context, the  objective function depends on the global society of players  (\textit{i.e.} agents). In other words, we will now unveil 
how the behaviors found in Section \ref{sct_exo} can also emerge from individual optimization strategies and why MFGs may actually be a natural mathematical framework to describe economic growth endogenously. In this section, we will show that the mean-field evolution encapsulated into Eq.(\ref{GENBURMO}) and the resulting propagating growth wave Eq.(\ref{PROLONG}) can alternatively be viewed as the ergodic solution of an associated MFG, in the sense of [Cardaliaguet et al., 2013]. While, in Section
\ref{sct_exo}, we had to study a forward-in-time problem, optimizing individual objective functions as to be done in this section generates a forward-/backward-in-time  structure which is typical for the MFG context. This reflects the underlying anticipation mechanism that animates the players' optimal decisions.

\subsection{Mean-Field Game and Sustained Growth}

\noindent We now focus on the cooperative parameter range $\eta \in ]- \infty ,2[$. The basic question to be addressed  is to construct an MFG that reproduces the flocking behavior calculated in Section 2.3. To construct this MFG, we are using the recent development exposed  in  \cite{SWIECICKI2016}, which is itself based on the MFG theory given in \textit{e.g.} \cite{LASRY2006A, GUEANT2011}.\\

\noindent We consider the class of  MFGs defined by:
\begin{equation}
\label{MFGBASE}
\left\{
\begin{array}{l}
dX_i(t) = a(X_i(t), t) dt + \sigma dW_i(t), \qquad \qquad i=1,,\cdots, N \\ \\
J\left(a(\cdot), X_i(\cdot)\right) = \mathbb{E} \left\{ 
 \int_{0}^{T} \underbrace{\left[  c\left(a(s), X_i(s)\right) - 
 V\left[\rho(\cdot,s),  X_i(s)\right)  \right]}_{\mathcal{L}(a(s), \rho(\cdot,s), X_i(s))} ds 
 + c_T(X_i(T))
 \right\}, 
\end{array}
\right.
\end{equation}
\noindent where $\mathbb{E} \left\{ \cdot \right\}$ is the average over the noise realizations, $\rho(x,t):= N^{-1} \sum_{i=1}^{i=N} \delta(x - X_i(t)) $ is the agents' empirical density, and $\mathcal{L}$ is an individual aggregated running cost due to innovation and imitation. When an agent increases her productivity, it leads not only to individual improved efficiency but also to a reduction in the number of peers she is in competition with. To take advantage of both aspects, there is an associated cost to be paid for these expected improvements. The time-dependent incurred cost $\mathcal{L}$ differs among the agents, depending on their productivity state, since it is easier for a laggard agent to improve than for one close to the technology frontier.\\

\noindent In the sequel, we choose:
\begin{equation}
\label{MFGBASE2}
\left\{
\begin{array}{l}
c\left(a(t), X_i(t)\right) = {\mu \over 2} \left[ (a(X_i(t), t)  - b\right]^{2}, \\ \\
V\left[ \rho(x,t), X_i(t) \right]= V[\rho(x,t)]= g \left[\rho(x,t) \right]^{p}, \quad g >0 \quad {\rm and }\quad p >0,
\end{array}
\right.
\end{equation}
\noindent where $g$ describes the wish for resemblance and hence the imitation activity of the agents, the parameter $p$ tunes the imitation strength, $\mu$ weights the drift adjustment cost, and $b$ denotes a target productivity growth rate. In a blind (\textit{i.e.} without interaction) and deterministic economy, each (isolated) agent would have to individually innovate at rate $b$. According to Eq.\eqref{MFGBASE2}, the interaction potential depends on the population density. This highlights that the growth process is not only due to innovators but also to the whole economy through the imitation mechanism. Note that except for the presence of the $b$ term, the objective function appearing in  Eqs.(\ref{MFGBASE}) and \eqref{MFGBASE2} coincides with the one given in \cite{SWIECICKI2016}.\\

\noindent By defining the value function
\begin{equation}
\label{VALUE}
u\left(x(t),t\right) := \min\limits_{a(\cdot)}\left\{ J\left(a(t),x(t)\right)\right\},
\end{equation}

\noindent the MFG reduces to solving the forward-/backward-in-time set of coupled PDEs:
\begin{equation}
\label{FORBACK}
\left\{
\begin{array}{l}
\partial_t \rho(x,t) = \partial_{x} \left[ \left(  {1 \over \mu}\partial_xu(x,t) -b  \right) \rho(x,t) \right] 
+ {\sigma^{2} \over 2} 
\partial^{2}_{xx}  \rho(x,t), \qquad {\rm (FP)}\\ \\ 
\partial_t u(x,t) -   \partial_x b\, u(x,t) - {1 \over 2\mu} \left[ \partial_x u(x,t) \right]^{2} +  {\sigma^{2} \over 2} 
\partial^{2}_{xx}  u(x,t) = -
g \left[\rho(x,t) \right]^{p}, \quad {\rm (HJB)}
\end{array}
\right.
\end{equation}
where the Hamilton-Jacobi-Bellman (HJB) equation describes the optimal control problem of each individual agent and the Fokker-Planck (FK) equation drives the evolution of the agent population. Observe that the $b$ component of the drift can be straightforwardly removed from Eq.(\ref{FORBACK})  by the Galilean transformation $ t \mapsto t'=t$ and $x \mapsto x':=(x - bt)$. This transformation of variables results in the following set of coupled PDEs:
\begin{equation}
\label{FORBACK2}
\left\{
\begin{array}{l}
\partial_t \rho(x,t) = \partial_{x} \left[ \left(  {1 \over \mu}\partial_xu(x,t)  \right) \rho(x,t) \right] 
+ {\sigma^{2} \over 2} 
\partial^{2}_{xx}  \rho(x,t), \qquad {\rm (FP)}\\ \\ 
\partial_t u(x,t)  - {1 \over 2\mu} \left[ \partial_x u(x,t) \right]^{2} +  {\sigma^{2} \over 2} 
\partial^{2}_{xx}  u(x,t) = -
g \left[\rho(x,t) \right]^{p}. \quad {\rm (HJB)}
\end{array}
\right.
\end{equation}
From this point, we follow the lines exposed in \cite{SWIECICKI2016} to get the resulting ergodic agent density, which takes the following form\footnote{For the convenience of the reader, the main steps of this calculation are shown briefly in Appendix D.}:

\begin{equation}
\label{ONDE}
\left\{ 
\begin{array}{l}
\rho(x) = {N \over \left[ \cosh \left(\beta x\right) \right]^{2/p }}, \\ \\ 
N = { \beta  \over  B\left({1\over 2}, {1 \over p} \right)},
\end{array}
\right.
\end{equation}

\noindent where $B\left({1\over 2}, {1 \over p} \right)$ is the Beta function (see \cite{GRADSHTEYN}, and the constant $\beta$ is given by Eq.(\ref{SOLSOL}) in Appendix D. Performing the inverse Galilean transformation, we obtain the following propagating soliton:
\begin{equation}
\label{ONDE1}
\rho(x-bt) = {N\over \left[ \cosh(\beta (x -bt)\right]^{2/p}}
\end{equation}

\noindent Proceeding to the following identifications: 
\begin{equation}
\label{IDENT}
\beta = 1,  \qquad  {2 \over p}= 2- \eta 
  \qquad {\rm and } \qquad b= \alpha + {1 \over 2} (2 - \eta) \sigma^{2} = \alpha + \frac{\sigma^{2}}{p},
\end{equation}
\noindent the stationary solution given by Eq.(\ref{PROLONG}) and  the ergodic state of  the MFG dynamics given by Eq.(\ref{ONDE1}) are identical.\\

\noindent Since, in Eq.(\ref{GENBURMO}),  the existence  of a  soliton is secured for the parameter range  $\eta \in [-\infty,2[$, it implies that a direct comparison with an MFG  exists only for $p \in [1 , \infty[$. Accordingly, for  $p \in [0,1]$ in  Eq.(\ref{IDENT}), there is no exogenous imitation strategy, as defined in Section 2, for which an ergodic state soliton that solves an MFG as defined by Eq (\ref{MFGBASE}) exists.\\

\noindent Solving the MFG determines the players' optimal trade-off between innovation and imitation in terms of costs. Ultimately, each agent individually optimizes her imitation patterns in order to reach the target productivity level $b$. A closer look at the specific form of Eq.(\ref{IDENT}) reveals that the parameter $p$ in Eq.(\ref{MFGBASE2}) affects only the gravity center  modulation strength $\eta$, and thus, the agents' imitation behavior. From Eq.(\ref{MFGBASE2}), we can transparently observe the complementary roles played by the individual cost $c\left(a(t), X_i(t)\right)$ and $V\left(\rho(x,t)\right) = g \left(\rho(x,t)^{p}\right)$. Specifically, we see that $V\left(\rho(x,t)\right)$ influences only the sharpness of the emerging soliton $\rho(x,t)$ given by Eq.(\ref{ONDE})) (\textit{i.e.} the swarm cohesion). Concerning the drift $b$, it also depends on the variance $\sigma^{2}$. From  Eq.(\ref{IDENT}), we emphasize that the sharpness of the soliton is directly correlated with the propagation speed. Specifically, wide solitons resulting from large values of $p$ travel with slower velocity compared with thin solitons obtained for smaller values of $p$. This is due to the stronger cooperation tendency implemented in the economy. As a result, the higher strength of mutual interaction between the players not only increases the growth rate but also reduces the inequality level. This is aligned with the conclusions of \cite{LUTTMER2012a}.


\section{Conclusion}\label{sct_conc}

\noindent At the interface between exact, life and social sciences, economic growth is an out-of static equilibrium  process which can be partly understood from a well balanced interplay between the firms' innovation capability which randomly drives the technological frontier, and imitation of the best ideas developed by technological leaders. This apparently banal ratchet mechanism should conceptually not be underestimated, since it is truly engrossing to realize how ubiquitous randomness may, if properly mastered, ultimately offer benefit. Perennial and relevant interdisciplinary models are necessarily based on stylization of simple and strongly universal underlying mechanisms.  It is worth realizing that idealizing economic growth by a systematic scavenging of the leaders' behavior is very generic, as the mathematical details of the driving noise and the imitation process do not influence the qualitative behavior. The idea that fluctuations may lead to benefit wipes off the intuitive idea that noise should be systematically filtered out, and economic growth is a perfect illustration of this natural paradigm. Developing economically relevant models which incorporate randomness (due to innovation) and intrinsic non-linearity (due to imitation), and for which one can explicitly keep track of the transient evolution, is an ongoing challenge. In this contribution, we propose a novel way to model how firms learn and imitate from the higher productivity leaders. By its intrinsic nature, imitation stimulates a mimetic  tendency (\textit{i.e.} an actual synchronization) which is permanently counterbalanced (or even destroyed) by the presence of noise. By modeling this subtle trade-off, we are able to explicitly show
how imitation mechanisms based on the proximity existing between concurrent firms in terms of productivity, actually play a central role in the innovation/imitation economic growth picture. Sustained growth, here stylized as stable propagating waves, can only emerge for strong enough imitation propensity.

\section{Acknowledgement}

The authors would like to thank the referees for their constructive comments, which enabled to strongly improve the present paper.


\bibliographystyle{apalike}

\bibliography{TITMOUSE}

\begin{thebibliography}{}

\bibitem[Acebr\'on et~al., 2005]{ACEBRON2005}
Acebr\'on, J.~A., Bonilla, L.~L., P\'erez~Vicente, C.~J., Ritort, F., and
  Spigler, R. (2005).
\newblock The kuramoto model: A simple paradigm for synchronization phenomena.
\newblock {\em Reviews of Modern Physics}, 77(1):137--185.

\bibitem[Achdou et~al., 2014]{ACHDOU2014}
Achdou, Y., Buera, F.~J., Lasry, J.-M., Lions, P.-L., and Moll, B. (2014).
\newblock Partial differential equation models in macroeconomics.
\newblock {\em Philosophical Transactions of the Royal Society A: Mathematical,
  Physical and Engineering Sciences}, 372:20130397.

\bibitem[Aghion and Howitt, 1992]{AGHION1992}
Aghion, P. and Howitt, P. (1992).
\newblock Model of growth through creative destruction.
\newblock {\em Econometrica}, 60(2):323--351.

\bibitem[Aghion and Howitt, 2009]{AGHION2009}
Aghion, P. and Howitt, P. (2009).
\newblock {\em The Economics of Growth}.
\newblock MIT Press.

\bibitem[Bedjaoui and LeFloch, 2004]{BEDJAOUI2004}
Bedjaoui, N. and LeFloch, P.~G. (2004).
\newblock Diffusive-dispersive traveling waves and kinetic relations versus
  singular diffusion and nonlinear dispersion.
\newblock {\em Proceedings of the Royal Society of Edinburgh}, 134(A):815--843.

\bibitem[Benhabib et~al., 2014]{BENHABIB2014}
Benhabib, J., Perla, J., and Tonetti, C. (2014).
\newblock Catch-up and fall-back through innovation and imitation.
\newblock {\em Journal of Economic Growth}, 19(1):1--35.

\bibitem[Benhabib et~al., 2017]{BENHABIB2017}
Benhabib, J., Perla, J., and Tonetti, C. (2017).
\newblock Reconciling models of diffusion and innovation: A theory of the
  productivity distribution and technology frontier.
\newblock {\em Econometrica}, R\&R.

\bibitem[Cardaliaguet et~al., 2013]{CARDALIAGUET2013}
Cardaliaguet, P., Lasry, J.-M., Lions, P.-L., and Porretta, A. (2013).
\newblock Long time average of mean field games with a nonlocal coupling.
\newblock {\em SIAM Journal on Control and Optimization}, 51(5):3558--3591.

\bibitem[Carmona and Delarue, 2018]{CARMONA2018}
Carmona, R. and Delarue, F. (2018).
\newblock {\em Probabilistic Theory of Mean Field Games with Applications}.
\newblock Springer.

\bibitem[Cucker and Smale, 2007]{CUCKER2007}
Cucker, F. and Smale, S. (2007).
\newblock Emergent behavior in flocks.
\newblock {\em IEEE Transactions on Automatic Control}, 52(5):852--862.

\bibitem[de~Geus, 2002]{DEGEUS2002}
de~Geus, A. (2002).
\newblock {\em The Living Company: Habits for Survival in a Turbulent Business
  Environment}.
\newblock Harvard Business School Press.

\bibitem[Debnath, 2005]{DEBNATH2005}
Debnath, L. (2005).
\newblock {\em Nonlinear Partial Differential Equations for Scientists and
  Engineers}.
\newblock Birkh\"auser, (second edition).

\bibitem[Dixit and Pindyck, 1994]{DIXIT1994}
Dixit, A.~K. and Pindyck, R.~S. (1994).
\newblock {\em Investment Under Uncertainty}.
\newblock Princeton University Press.

\bibitem[Gomes et~al., 2015]{GOMES2015}
Gomes, D.~A., Nurbepkyan, L., and Pimental, E.~A. (2015).
\newblock {\em Economic Models and Mean-Field Games}.
\newblock Publicacaoes Mathematics.

\bibitem[Gradshteyn and Ryzhik, 1980]{GRADSHTEYN1980}
Gradshteyn, I.~S. and Ryzhik, M. (1980).
\newblock {\em Tables of Integrals, Series and Products}.
\newblock Academic Press.

\bibitem[Grossman and Helpman, 1991]{GROSSMAN1991}
Grossman, G.~M. and Helpman, E. (1991).
\newblock Quality ladders in the theory of growth.
\newblock {\em The Review of Economic Studies}, 58(1):43--61.

\bibitem[Gu\'eant, 2012]{GUEANT2012}
Gu\'eant, O. (2012).
\newblock Mean field games equations with quadratic hamiltonian: A specific
  approach.
\newblock {\em Mathematical Models and Methods in Applied Sciences}, 22(9).

\bibitem[Gu\'eant et~al., 2011]{GUEANT2011}
Gu\'eant, O., Lasry, J.~M., and Lions, P.~L. (2011).
\newblock Mean field games and applications.
\newblock 2003:205--266.

\bibitem[Henkin and Polterovich, 1991]{HENKIN1991}
Henkin, G.~M. and Polterovich, V.~M. (1991).
\newblock Schumpeterian dynamics as a non-linear wave theory.
\newblock {\em Journal of Mathematical Economics}, 20(6):551--590.

\bibitem[Hongler et~al., 2014]{HONGLER2014}
Hongler, M.-O., Filliger, R., and Gallay, O. (2014).
\newblock Local versus nonlocal barycentric interactions in 1d dynamics.
\newblock {\em Mathematical Bioscience and Engineering}, 11(2):323--351.

\bibitem[Hopenhayn, 1992]{HOPENHAYN1992}
Hopenhayn, H.~A. (1992).
\newblock Entry, exit, and firm dynamics in long run equilibrium.
\newblock {\em Econometrica}, 60(5):1127--1150.

\bibitem[Ichiba et~al., 2011]{ICHIBA2011}
Ichiba, T., Papathanakos, V., Banner, A., Karatzas, I., and Fernholz, R.
  (2011).
\newblock Hybrid atlas models.
\newblock {\em Annals of Applied Probability}, 21(2):609--644.

\bibitem[Iwai, 1984]{IWAI1984}
Iwai, K. (1984).
\newblock Schumpeterian dynamics - an evolutionary model of innovation and
  imitation.
\newblock {\em Journal of Economic Behavior and Organization}, 5(2):159--190.

\bibitem[Iwai, 2000]{IWAI2000}
Iwai, K. (2000).
\newblock A contribution to the evolutionary theory of innovation, imitation
  and growth.
\newblock {\em Journal of Economic Behavior and Organization}, 43(2):167--198.

\bibitem[Koenig et~al., 2016]{KOENIG2016}
Koenig, M.~D., Lorenz, J., and Zilibotti, F. (2016).
\newblock Innovation vs. imitation and the evolution of productivity
  distributions.
\newblock {\em Theoretical Economics}, accepted for publication.

\bibitem[Lachapelle et~al., 2010]{LACHAPELLE2010}
Lachapelle, A., Salomon, J., and Turinici, G. (2010).
\newblock Computation of mean field equilibria in economics.
\newblock {\em Mathematical Models and Methods in Applied Sciences},
  20(4):567--588.

\bibitem[Lasry and Lions, 2006a]{LASRY2006A}
Lasry, J.-M. and Lions, P.-L. (2006a).
\newblock Mean field games - finite horizon and optimal control.
\newblock {\em Comptes Rendus Math\'ematiques}, 343(10):679--684.

\bibitem[Lasry and Lions, 2006b]{LASRY2006B}
Lasry, J.-M. and Lions, P.-L. (2006b).
\newblock Mean field games - the stationary case.
\newblock {\em Comptes Rendus Math\'ematiques}, 343(9):619--625.

\bibitem[Lasry and Lions, 2007]{LASRY2007}
Lasry, J.-M. and Lions, P.-L. (2007).
\newblock Mean field games.
\newblock {\em Japanese Journal of Mathematics}, 2(1):229--260.

\bibitem[Lucas~Jr., 2009]{LUCAS2009}
Lucas~Jr., R.~E. (2009).
\newblock Ideas and growth.
\newblock {\em Economica}, 76(301):1--19.

\bibitem[Lucas~Jr. and Moll, 2014]{LUCAS2014}
Lucas~Jr., R.~E. and Moll, B. (2014).
\newblock Knowledge growth and the allocation of time.
\newblock {\em Journal of Political Economy}, 122(1):1--51.

\bibitem[Luttmer, 2007]{LUTTMER2007}
Luttmer, E. G.~J. (2007).
\newblock Selection, growth, and the size distribution of firms.
\newblock {\em The Quarterly Journal of Economics}, 122(3):1103--1144.

\bibitem[Luttmer, 2012a]{LUTTMER2012a}
Luttmer, E. G.~J. (2012a).
\newblock Eventually, noise and imitation implies balanced growth.
\newblock {\em Federal Reserve Bank Minneapolis}, Working Paper 699.

\bibitem[Luttmer, 2012b]{LUTTMER2012b}
Luttmer, E. G.~J. (2012b).
\newblock Technology diffusion and growth.
\newblock {\em Journal of Economic Theory}, 147:602--622.

\bibitem[Montroll and Lebowitz, 1979]{MONTROLL1979}
Montroll, E.~W. and Lebowitz, J.~L. (1979).
\newblock {\em Studies in Statistical Mechanics}.
\newblock North Holland.

\bibitem[Nelson and Winter, 1982]{NELSON1982}
Nelson, R. and Winter, S. (1982).
\newblock {\em An Evolutionary Theory of Economic Change}.
\newblock Belknap Press/Harvard University Press.

\bibitem[Nicolis and Prigogine, 1992]{NICOLIS1989}
Nicolis, G. and Prigogine, I. (1992).
\newblock {\em Exploring Complexity}.
\newblock Freeman, New-York.

\bibitem[Reinganum, 1985]{REINGANUM1985}
Reinganum, J.~F. (1985).
\newblock Innovation and industry evolution.
\newblock {\em The Quarterly Journal of Economics}, 100(1):81--99.

\bibitem[Ridley, 2010]{RIDLEY2010}
Ridley, M. (2010).
\newblock {\em The Rational Optimist. How Prosperity Evolves}.
\newblock Harper-Collins, NY.

\bibitem[Romer, 1990]{ROMER1990}
Romer, P. (1990).
\newblock Endogenous technological change.
\newblock {\em Journal of Political Economy}, 98(5):1002--1037.

\bibitem[Saviotti and Mani, 1998]{SAVIOTTI1998}
Saviotti, P.~P. and Mani, G.~S. (1998).
\newblock Technological evolution, self-organisation and knowledge.
\newblock {\em The Journal of High Technology Management Research},
  9(2):255--270.

\bibitem[Staley, 2011]{STALEY2011}
Staley, M. (2011).
\newblock Growth and the diffusion of ideas.
\newblock {\em Journal of Mathematical Economics}, 47(4-5):470--478.

\bibitem[Swiecicki et~al., 2016]{SWIECICKI2016}
Swiecicki, I., Gobron, T., and Ullmo, D. (2016).
\newblock Schroedinger approach to mean field games.
\newblock {\em Physical Review Letters}, 116(12).

\bibitem[Wyles et~al., 1983]{WYLES1983}
Wyles, J.~S., Kunkel, J.~G., and Wilson, A.~C. (1983).
\newblock Birds, behavior, and anatomical evolution.
\newblock {\em Proceedings of the National Academy of Sciences of the United
  States of America}, 80(14 I):4394--4397.

\bibitem[Yin et~al., 2012]{YIN2012}
Yin, H., Mehta, P.~G., Meyn, S.~P., and Shanbhag, U.~V. (2012).
\newblock Synchronisation of coupled oscillators is a game.
\newblock {\em IEEE Transactions on Automatic Control}, 57(4):920--935.

\end{thebibliography}


\newpage
\section*{Appendix A}

\noindent {\bf List of the principal variables and notations.}

\begin{itemize}
    \item[$\bullet$] ${\cal A}_k$: $k^{{\rm th}}$ firm (equivalently also called agent)  forming the economy of $N$ entities.
    \item[$\bullet$]$A_k(t)$: positive definite stochastic process  with state space either $\mathbb{N}$ (in the Markov chain description)  or  $\mathbb{R}^{+}$ (in the diffusion process description). It describes the $t$-dependent position of  the total factor productivity of the $k^{{\rm th}}$ firm at time $t$.
    \item[$\bullet$]$\vec{A}(t)= \left( A_1(t), A_2(t), \cdots, A_N(t) \right)$: vector formed with the $N$ individual firms'  total factor productivity.
    \item[$\bullet$] $A(t)$: a representative (i.e., randomly chosen component) of the vector process $\vec{A}(t)$.
    \item[$\bullet$]  $a( X_k(t), t)$: general drift of the diffusion process entering into  the MFG description.
    \item[$\bullet$] $\alpha$:   constant drift  component  of   the  $X_k(t)$-dynamics for $k=1,2,\cdots, N$; this drift component is common to all firms.
    \item[$\bullet$] $\Delta$:  size of a discrete time step  in the discrete time Markov chain describing the random  evolution on a Schumpeterian ladder representing factor  productivity .
    \item[$\bullet$] ${\mathcal D}_k ( \vec{X}(\nu \Delta)$ : imitation jump probability bias for the $k^{{\rm th}}$ firm, $k=1,...,N$, at the discrete time $\nu \Delta$. This bias is induced by the mutual interactions among the $N$ Markov chains describing the economy.
     \item[$\bullet$] $G(x,t) = \int_{x}^{\infty} \rho(y,t) dy $: complementary distribution function.
    \item[$\bullet$] $\hat{G}(x,t) = \int_{- \infty}^{x} \rho(y,t) dy $: distribution function.
    \item[$\bullet$] $\gamma$: control  parameter tuning the imitation drift sensitivity due to the  presence of  leaders.
    \item[$\bullet$] $\Gamma$:   shorthand notation which, depending on the context, stands  either for ${\gamma U \over 2}$ or for ${\gamma \over 2}$ (see Eq.\eqref{GAMMA}).
    \item[$\bullet$] ${\mathcal G} (x- \langle X(t) \rangle ), \mathbb{R} \rightarrow \mathbb{R}^{+}$: barycentric weight factor.
    \item[$\bullet$]${\mathcal J}\left(X_k(t), \rho(x,t)  \right)$: interaction kernel describing the MFG imitation process of firm ${\mathcal A}_k$.
    \item[$\bullet$] $ka $: position of the $k^{{\rm th}}$ rung on a Schumpeterian ladder representing factor productivity.
    \item[$\bullet$] ${\mathcal N}_k (t)\in [0, N] $: number of leaders ahead of firm ${\mathcal A}_k$ at time $t$ on the Schumpeterian ladder representing factor productivity.
    \item[$\bullet$] $P( ka , \nu \Delta)$: probability to occupy  the $k^{{\rm th}}$ rung  at  time $\nu \Delta$   on the  Schumpeterian ladder representing factor productivity. $P( ka , \nu \Delta)$ is the solution of  the master equation   associated with the Schumpeterian  ladder Markov chain.
    \item[$\bullet$] $p( ka , \nu \Delta, \vec{X}(\nu\Delta) $:  probability  of a  single rung jump upward on the discrete time Markov chain describing the evolution on the Schumpeterian ladder representing factor productivity.
    \item[$\bullet$]$ \varphi(x,t)$: shorthand notation which, depending on the context, stands  either for $\rho(x,t)$ or for  $G(x,t)$ (see Eq.\eqref{BURGO}).
    \item[$\bullet$] $q( ka , \nu \Delta, \vec{X}(\nu\Delta) $:  probability  of a  single rung jump downward in the discrete time Markov chain describing the evolution on the Schumpeterian ladder representing the factor of  productivity.
    \item[$\bullet$]$ \rho(x,t)dx:= {\rm Prob} \left\{  x \leq X(t)\leq x+dx\right\} $: (normalized) probability density of the stochastic process $X(t)$.
    \item[$\bullet$] $\sigma dW(t)$: White Gaussian Noise process with variance $\sigma^{2}$.
    \item[$\bullet$] $u(x,t)$: value function solving  the Hamilton-Bellman-Jacobi problem resulting from the MFG approach.
    \item[$\bullet$]$U \in [0, \infty]$: size of the observation window within which firms count the number of their  productivity leaders.
    \item[$\bullet$] $V(\rho(x,t)) = g \rho^{p}(x,t)$: imitation running cost in the MFG description ($g, p \in \mathbb{R}^{+}$ are both  constants).
    \item[$\bullet$]$X_k(t) := \ln (A_k(t))$:  stochastic process with state space $\mathbb{Z}$ (in the Markov chain description) or $\mathbb{R}$ (in the diffusion process description). It describes the $t$-dependent position of the logarithm of the total factor productivity of the $k^{{\rm th}}$ firm at time $t$.
    \item[$\bullet$] $\vec{X}(t) = \left( X_1(t), X_2(t), \cdots, X_N(t)\right) $: vector formed with the $N$ individual firms'  logarithm of their total factor productivity.
    \item[$\bullet$] $X(t)$: a representative (i.e., a randomly chosen) component  of the vector process $\vec{X}(t)$.
    \item[$\bullet$] $\langle X^{n} (t) \rangle  \equiv  \mathbb{E} \left\{ X^{n}(t)  \right\}  := \int_{\mathbb{R} } x^{n} \rho(x,t) dx$.
\end{itemize}


\section*{Appendix B}

\noindent {\bf Titmice versus robins: How territorial imitation ranges drastically affect collective dynamics.} \\ 

\noindent To illustrate the potential role played by the observation range in imitation processes, let us turn toward ornithology and consider a situation originally studied by J. S. Wyles et al. in \cite{WYLES1983}). The authors develop the idea that evolution is essentially driven by species behavior, rather than by the environment only. To support this view, they consider the behavior of songbirds in Great Britain. According to \cite{WYLES1983}, at the beginning of the 20th century, British milkmen used to leave milk bottles without caps outside people's homes. Two species of songbirds, the titmouse and the robin, learned to feed on cream from these milk bottles. Then came an innovation in the milk industry in the 1930s: covering milk bottles with aluminum bottle seals. According to \cite{WYLES1983}, the titmouse learned to pierce the aluminum seals and, in a matter of two decades, successfully spread this newly acquired technique across their entire species throughout all of Great Britain, estimated at the time to be about a million individuals. In contrast, the robin never widely learned the technique for drilling through the aluminum seals. The reason behind this was the robin's territorial inclination and the relative isolation of individuals, which inhibited the spread of innovation. Further investigation showed that the titmouse is mobile, and its behavior promoted the propagation of the new approach. Because robins mainly act alone, they lacked the capability of exploring new opportunities that existed  in their environment. In contrast, the non-territorial titmouse is able to learn and adapt to its environment in a quick and agile manner thanks to its natural and efficient group mobility. In \cite{DEGEUS2002}, A. de Geus exhibits a parallel between this ornithological example and the capacity to learn and to adapt quickly in economic environments. De Geus shows that these aspects are determinant features in the long-term survival of companies.


\section*{Appendix C}

\noindent {\bf Solving the dynamics in the presence of the conformism modulation factor.}\\ 

\vspace{-0.05cm}
\noindent As in Section 2, we are interested in  the possibility of observing  a constant variance  stable  wave  of the form $\rho(x- (\alpha + w) t): = \rho(\xi)$  traveling with constant velocity $(\alpha + w)$, for the dynamics given by Eq.(\ref{GENBURMO}). As a function of  the new variable $\xi =\left[ x-(\alpha + w)t\right]$, Eq.(\ref{GENBURMO}) takes the form:
\begin{equation}
\label{BUSTA}
0 = \partial_{\xi}\left[ \rho(\xi)\left\{ w-  \int_{\xi}^{\infty} {\cal G}\left(z  \right)\rho(z)  dz \right\}  + {\sigma^{2} \over 2} \partial_{\xi} \rho(\xi)
\right].
\end{equation}
As the wave is assumed to be stationary with constant variance and  with traveling velocity $(\alpha + w)$, this imposes an additional constraint  that $w$ has to satisfy, namely: 
\begin{equation}
\label{Condition}
\int_{\mathbb{R}} \xi \rho (\xi) d\xi =0.
\end{equation}
Integrating Eq. \eqref{BUSTA} once with respect to  $\xi$ (with zero integration constant, as no probability current is sustained in the stationary regime), we get:

\begin{equation}
\label{ONEINTERGRA}
0 = \rho(\xi)\left\{ w-  \int_{\xi}^{\infty} {\cal G}\left(z  \right)\rho(z)  dz \right\}  + {\sigma^{2} \over 2} \partial_{\xi} \rho(\xi).
\end{equation}

\noindent Then dividing by $\rho(\xi)>0$, Eq.(\ref{ONEINTERGRA}) can be rewritten as:
\begin{equation}
\label{COMPAT}
-{\sigma^{2} \over 2} \, \partial_{\xi} \log \left[ \rho(\xi) \right] = \left\{ w-  \int_{\xi}^{\infty} {\cal G} \left(z \right)\rho(z)  dz \right\}.
\end{equation}\\ 
 {\bf Modulating $\mathcal{G}$ factor for which an exact analytic resolution is possible}.\\

\begin{figure}[h]
\begin{center}
\hspace{-0.4cm}
\includegraphics[width=115mm, height=80mm]{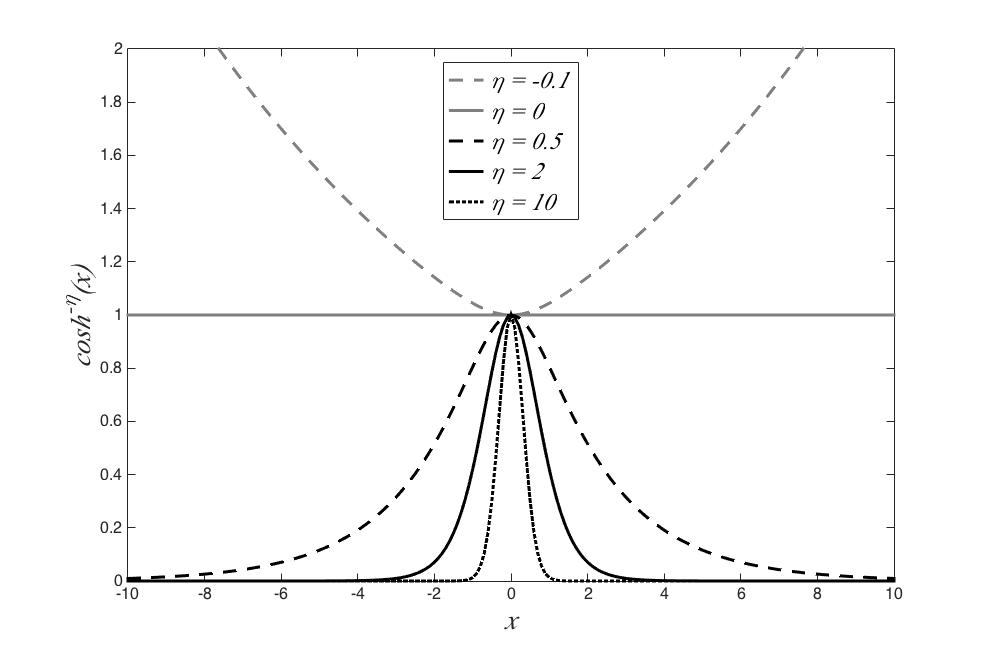}
\end{center}
\caption{Considered class of barycentric modulation functions $\cosh^{-\eta}(x)$. The transition between propagation regimes emerges at $ \eta = 2$.}
\label{cosinush}
\end{figure}

\vspace{-0.05cm} 
\noindent Solving the nonlinear integro-differential equation given by Eq.(\ref{COMPAT}) exactly is  not feasible in general.  However, this is possible for symmetric barycentric modulation functions ${\cal G}$ of the form (see Figure \ref{cosinush} for an illustration):
\begin{equation}
\label{MODULO}
{\cal G}(x) = {\cal A} \cosh^{-\eta}(x), \qquad \eta \in \mathbb{R}^{+},
\end{equation}
with ${\mathcal A}>0 \in \mathbb{R}^{+}$  and $\eta \in \mathbb{R}^{+}$.

\noindent We now verify that  it exists a solution of  Eq.(\ref{COMPAT}) in the form:
\begin{equation}
\label{ANSATZ}
\rho(\xi) = \mathcal{N}(m) \cosh^{-m}(\xi) \qquad m \in \mathbb{R}^{+},
\end{equation}
where  $\mathcal{N}(m)$ is a normalization factor ensuring that $\int_{\mathbb{R}}  \mathcal{N}(m) \cosh^{m}(\xi) d\xi=1$. Namely, here, ${\cal N}(m) = \Gamma[(m+1)/2] / \sqrt{\pi}\, \Gamma(m/2)$, with $\Gamma(z)$ standing for the gamma function.  Note that the existence of $\mathcal{N}(m) $ is ensured for  $m<0$. The choice given in Eq.(\ref{ANSATZ}) implies that $\rho(\xi) = \rho(-\xi)$, and therefore, Eq.(\ref{Condition}) is automatically satisfied. Plugging Eq.(\ref{ANSATZ})  into Eq.(\ref{COMPAT}), we find:
\begin{equation}
\label{CALCUL}
{\sigma^{2} \over 2}\,  m\tanh(\xi) =  \left[w-      {\cal A} \mathcal{N}(m)\int_{\xi}^{\infty} \cosh^{-(\eta + m)}(\xi)  d\xi \right].
\end{equation}
Using the identity  $\int_{x}^{\infty} \cosh(x)^{-2} dx = \left[ 1-\tanh(x)\right]$, we verify that Eq.(\ref{CALCUL}) is exactly solved, provided that we  simultaneously  impose that:

$$
\eta + m = 2, \qquad { \mathcal A}= { \sigma^{2} m\over 2 \mathcal{N}(m) } \qquad {\rm and} \qquad w ={\mathcal A}  \mathcal{N}(m), 
$$

\noindent which can be rewritten as:

\begin{equation}
\label{SOLBU}
m = 2 - \eta , \qquad {\cal A} = { \sigma^{2} \sqrt{\pi} m  \Gamma(m/2)\over 2  \Gamma[(m+1)/2]}  =  {(2- \eta) [\Gamma(1 - {\eta \over 2} ]^{2}\over 2^{\eta}\Gamma(2- \eta)}\, \sigma^{2}, \qquad w = {1 \over 2} m \sigma^{2}.
\end{equation}

\noindent From Eqs.(\ref{CALCUL}) and (\ref{SOLBU}), we conclude that  for  $\eta \in ]-\infty,2[ \Rightarrow m > 0$ ({\it i.e.}  slow decay of the ${\cal G}(x)$ modulation, leading to longer-range interactions), ${\cal N}(m)$ exists, and a stable traveling solitary wave with velocity $w$ is created.  Conversely, when $\eta \in [2, \infty[ \Rightarrow m < 0$ ({\it i.e.} rapid decay of the ${\cal G}(x)$ modulation, leading to short-range interactions), Eq.(\ref{SOLBU})  collapses as  ${\cal A}<0$,  and $\rho(\xi)$ in Eq.(\ref{ANSATZ}) is not normalizable. In this case, no stable solitary wave can be sustained for this $\eta$-parameter range.


\section*{Appendix D}

After performing the Galilean transformation of variables (we omit the primes for notation convenience), we proceed as shown in \cite{SWIECICKI2016}. Accordingly, we introduce the Hopf-Cole logarithmic transformation   $u(x,t) =  - \mu \sigma^{2} \ln\left[ \Phi(x,t) \right]$ and $\Gamma(x,t) = m(x,t)/ \Phi(x,t)$ \cite{SWIECICKI2016,GUEANT2012}), and the coupled PDEs  Eq.(\ref{FORBACK}) transform to the set of nonlinear Schr\"odinger (NLS)-like equations:

\begin{equation}
\label{SCHROEDINGER}
\left\{
\begin{array}{l}

- \mu \sigma^{2} \partial_t\Phi(x,t) = {\mu \sigma^{4} \over 2} \partial_{xx} \Phi(x,t) + V[\rho(x,t)] \Phi(x,t), \\ \\
 + \mu \sigma^{2} \partial_t \Gamma(x,t) = {\mu \sigma^{4} \over 2} \partial_{xx} \Gamma(x,t) + V[\rho(x,t)] \Gamma(x,t).

\end{array}
\right.
\end{equation}

\noindent Invoking, as in  \cite{SWIECICKI2016}, the fundamental contribution of  \cite{CARDALIAGUET2013}, we focus on  times $0<< t << T$, for which  the dynamics is essentially insensitive to the  boundary conditions. Focusing on this quasi-stationary ergodic state,  as in  \cite{SWIECICKI2016}, we write  $\Phi(x,t) = e^{-{\epsilon \over \mu \sigma^{2}}t} \Psi(x)$ and $\Gamma(x,t) = e^{{\epsilon \over \mu \sigma^{2}}t} \Psi(x)$ which leads to the NLS equation:

\begin{equation}
\label{ONDE}
{\mu \sigma^{4} \over 2} \partial_{xx} \Psi(x) + V[\rho(x)] \Psi(x)  = \epsilon \Psi(x) .
\end{equation}

\noindent Now, we observe that with the specific choice:
\begin{equation}
\label{VSPECIFIC}
V[\rho(x)]= g\left[ \rho(x) \right]^{p} = g \left[\Psi(x) \right]^{2p}, \qquad g>0, 
\end{equation}

\noindent  Eq.(\ref{SCHROEDINGER}) can be integrated by separation of variables, namely:

\begin{equation}
\label{INTEGRO}
dx={
d\Psi(x) \over \sqrt
{
{2\epsilon \over \mu \sigma^{4} } \Psi(x)^{2} 
- {2 g \over (p+1) \mu \sigma^{4}} \Psi(x)^{2(p+1)}
}
},
\end{equation}

\noindent  Using the identity $\cosh^{2}(z) -1 = \sinh^{2}(z)$, we can directly verify that Eq.(\ref{INTEGRO}) is solved by  the soliton-like (normalized) wave function:

\begin{equation}
\label{SOLSOL}
\left\{
\begin{array}{l}
\Psi(x) =  {\sqrt{N} \over \left[\cosh(\beta x)\right]^{1/p}},\qquad \left( \int_{\mathbb{R}} {dx \over \Psi^{2}(x) }=1\right),\\ \\
\beta= {p \sqrt{2 g  N^{p}} p\over(p+1) \sqrt{\mu} \sigma^{2}}, \\ \\
N= {\beta\over B\left[ {1\over 2}, {1\over p}\right]}, \\ \\
\epsilon ={gN^{p} \over (p+1) }.
\end{array}
\right.
\end{equation}

\noindent where $B(x,y):= {\Gamma(x) \Gamma(y)\over \Gamma(x+y)}$ stands for the  Beta function. In particular,  we may obtain $\beta =1$ for  an appropriate choice of the MFG control parameters $\mu, g, p, \sigma$. Finally, the  ergodic agents density itself follows directly as $\rho(x) = \left[\Psi(x) \right]^{2}$.


\section*{Appendix E}\label{sct_sim}

\begin{figure}[H]
\begin{center}
\hspace{-0.2cm}
\includegraphics[width=140mm, height=160mm]{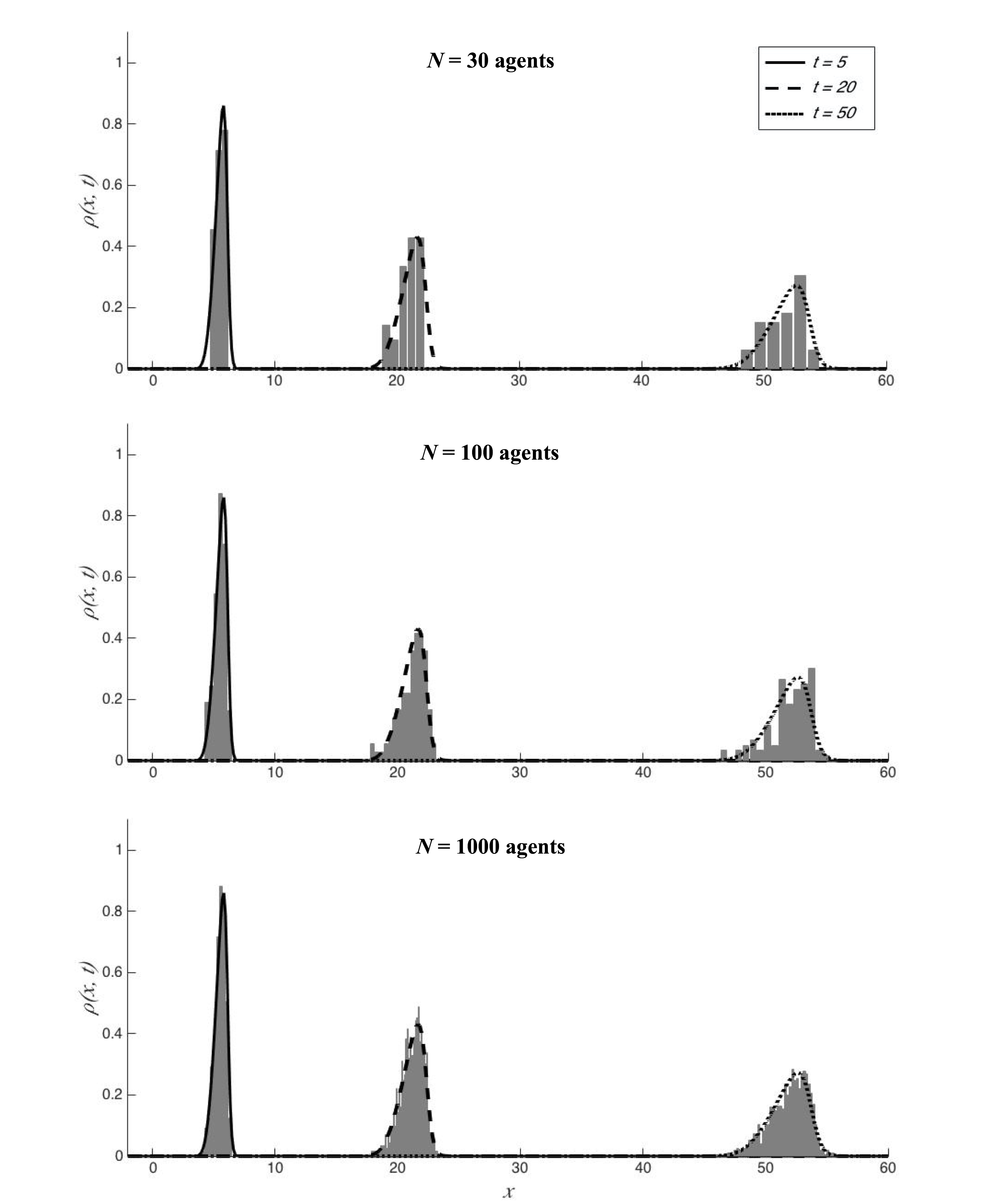}
\caption{Simulated collective dynamics observed for the infinitesimal imitation range when $U=0.1$, $\gamma=1$, $\alpha=1$, $\sigma=0.2$, and time discretization $\Delta t = 0.1$. The simulated histograms confirm the absence of a stable  growing productivity wave, as predicted by Eq.\eqref{GENOBURO}.}
\label{graph_burger_total}
\end{center}
\end{figure}

\begin{figure}[H]
\begin{center}
\hspace{-0.2cm}
\includegraphics[width=140mm, height=150mm]{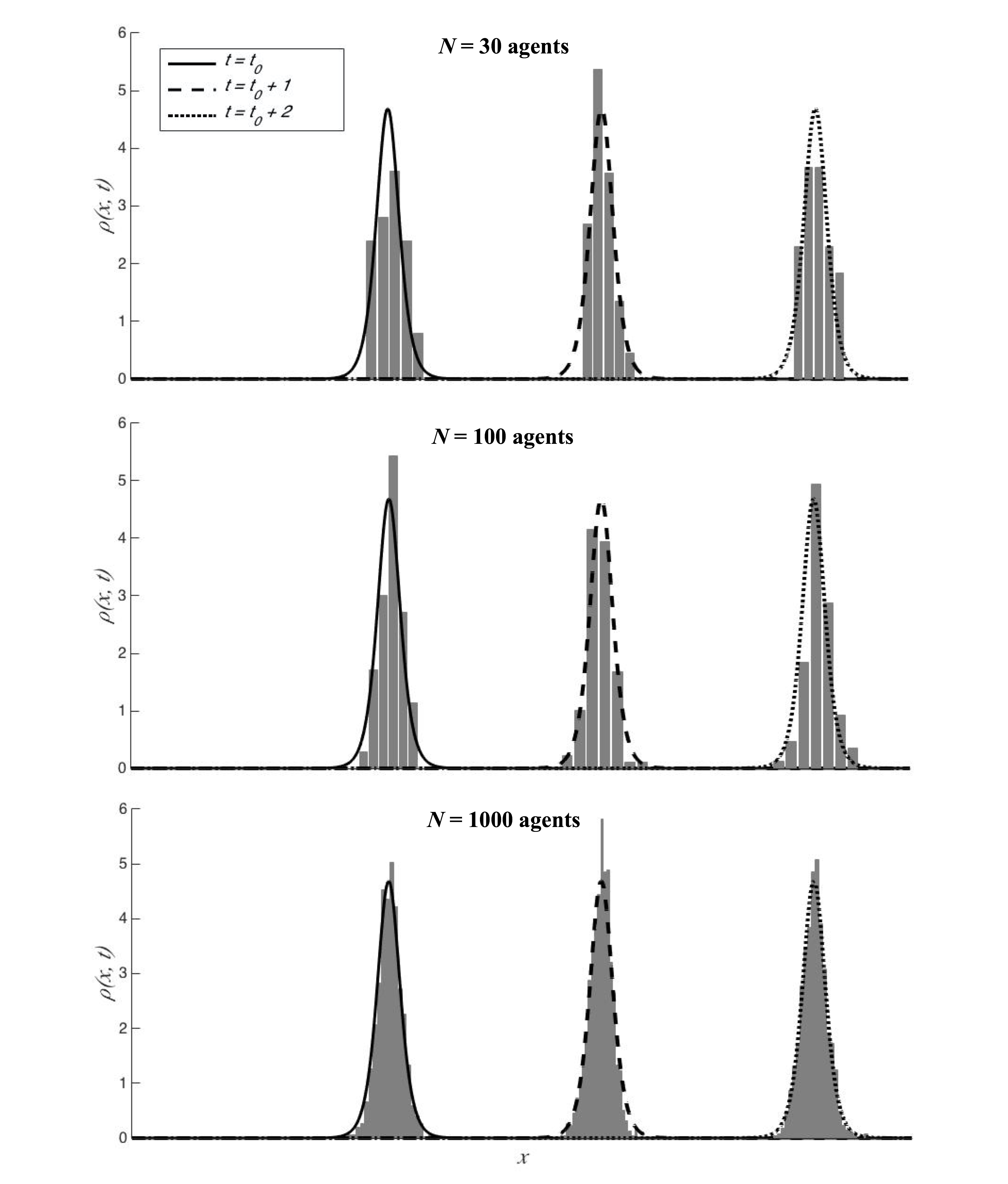}
\caption{Simulated collective dynamics observed for the infinite imitation range when $U=1000$, $\gamma=1$, $\alpha=1$, $\sigma=0.2$, and time discretization $\Delta t = 0.1$. The simulated histograms confirm the generation of a stable  growing productivity wave with constant variance, as predicted by Eq.\eqref{ADENS}.}
\label{graph_travelling_total}
\end{center}
\end{figure}

\begin{itemize}
\item[\textit{(a)}] \noindent {\bf Finite Population of Agents}\\

\vspace{-0.45cm}
\noindent Strictly speaking, the mean-field approximation made in Eq.\eqref{CLIMIT} requires an infinite number $N$ of agents to provide exact results. Nevertheless, as highlighted by the simulation results displayed in Figures \ref{graph_burger_total} and \ref{graph_travelling_total}, the mean-field population dynamics given by Eqs.\eqref{GENOBURO} and \eqref{ADENS} is already observed for the limited population of agents $N= 30, 100$, and $1000$.
\end{itemize}

\noindent Table \ref{table_finite_number_agents} provides a characterization of the theoretical and simulated distributions obtained for the stable growing productivity regime displayed in Figure \ref{graph_travelling_total}. In accordance with \cite{ROMER1990}, we observe from simulation experiments that the growth rate displays a small rise when the size of the economy increases. Conversely, no clear tendency is observed with respect to the variance and the kurtosis of the firms' productivity, hence suggesting that a rather small economy (i.e., more than $30$ economical agents) will already showcase the general behavior described by our model. Simulation experiments show however a tendency for the skewness to diminish as the size of the economy increases. This suggests that a critical number of economical agents is required to ensure the creation of a stable growing productivity regime.

\begin{table}[htbp]
\begin{center}
\caption{Characterization of the theoretical and simulated distributions obtained for the stable growing productivity regime displayed in Figure \ref{graph_travelling_total} for $t=t_0 +2$ .}
\label{table_finite_number_agents}
\renewcommand{\tabcolsep}{6pt}
{\scriptsize
\renewcommand{\arraystretch}{1.5}
\vspace{0.3cm}
\begin{tabular}{|c|c|c|c|c|}
\hline
&$N=30$ & $N=100$ & $N=1000$ & $N=\infty$\\
\hline
\hline
\hline
Mean & 32.4641 & 32.6269 & 32.8130 & 32.9936 \\
\hline
Variance & 0.0084 & 0.0099 & 0.0079 & 0.0095\\
\hline
Skewness & 0.3153 & -0.1337 & 0.1111 & 0  \\
\hline
Kurtosis & 2.2104 & 4.0779 & 3.3982 & 4.1762\\
\hline
\end{tabular}
}
\end{center}
\end{table}

\begin{figure}[H]
\begin{center}
\hspace{-0.2cm}
\includegraphics[width=140mm, height=50mm]{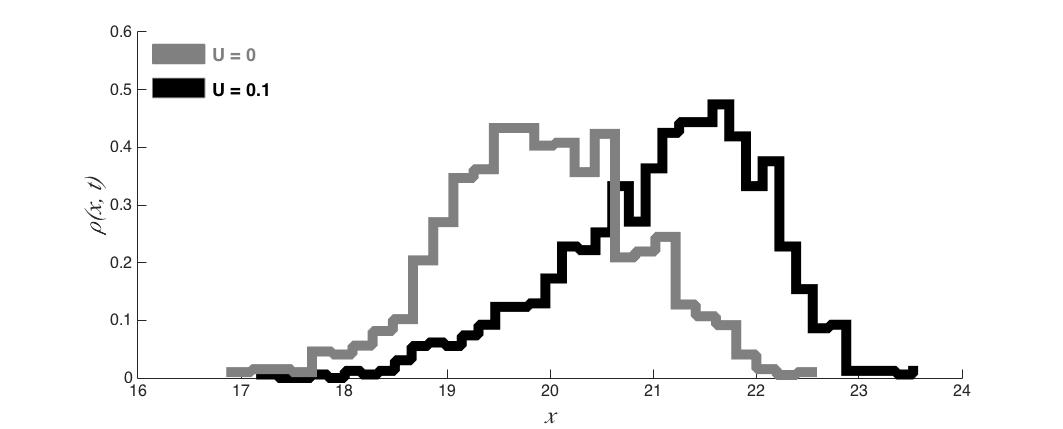}
\caption{Simulated collective dynamics observed for $N=1000$ agents at $t=20$, when $\gamma=1$, $\alpha=1$, $\sigma=0.2$, and time discretization $\Delta t = 0.1$. When the imitation range $U=0$, purely diffusive behavior is observed, where $\rho(x,t)$ is symmetric and propagates at speed $\alpha$. When $U=0.1$, as predicted by Eq.\eqref{GENOBURO}, the interactions between the agents produce an asymmetric shape for the density $\rho(x,t)$ and cause the propagation speed to be equal to $(\alpha + \gamma U/2)$.}
\label{graph_U_comparison}
\end{center}
\end{figure}

\vspace{1.85cm}
\begin{itemize}
\item[\textit{(b)}] \noindent {\bf Arbitrary Interaction Range $\boldsymbol{U}$.}\\

\vspace{-0.35cm}
To appreciate the influence of short interaction ranges as considered in Eq.\eqref{GENOBURO}, the situation with $U=0.1$ is compared to the strictly myopic situation that arises when $U=0$ (\textit{i.e.} strictly independent agents evolve as $N$ constant drifted Brownian motions). As shown in Figure \ref{graph_U_comparison}, the purely diffusive behavior obtained for $U=0$ noticeably differs from the interactive dynamics even for a small $U$.
\end{itemize}

\vspace{-0.6cm}
\begin{itemize}
\item[] When the imitation range $U$ lies in-between the two limiting regimes solved in Eqs.\eqref{GENOBURO} and \eqref{ADENS}, the propagation speed of the agent population density $\rho(x,t)$ is observed  in Figure \ref{graph_U_variation} to monotonously increase with $U$. Augmenting the interaction range $U$ enhances the average traveling velocity of the whole population. Specifically, the propagation speed is due to two contributions: 
\begin{figure}[h]
\begin{center}
\hspace{-0.2cm}
\includegraphics[width=140mm, height=70mm]{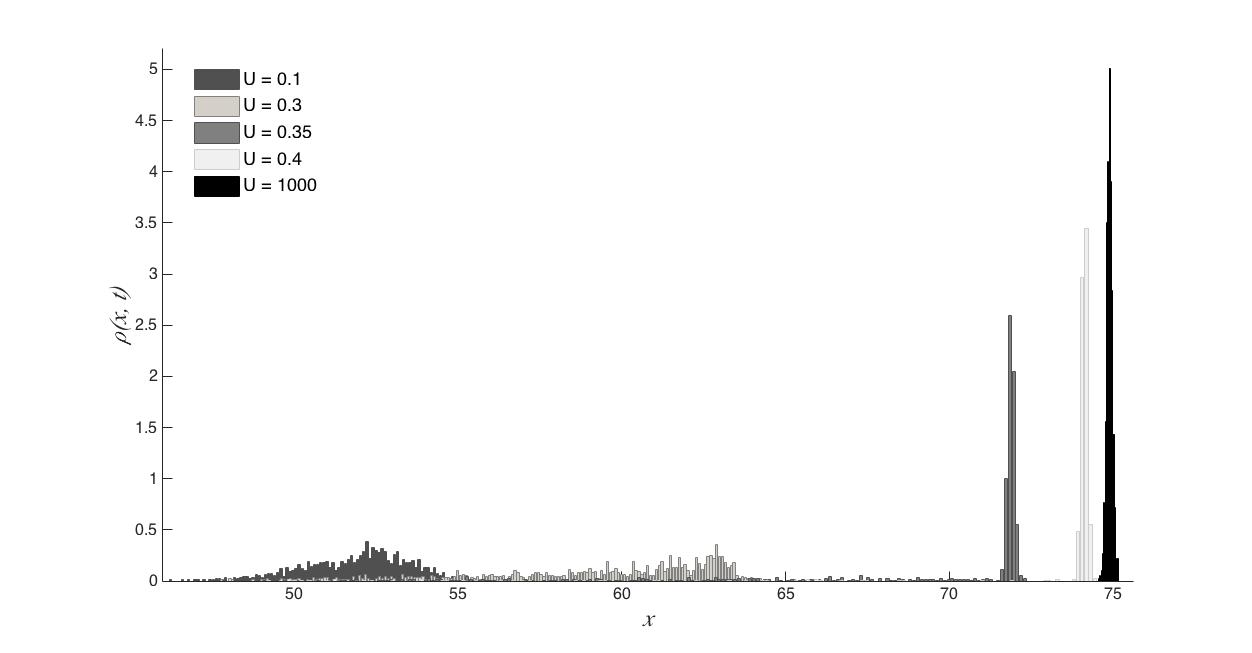}
\caption{Simulated collective dynamics observed for $N=1000$ agents at $t=50$, when $\gamma=1$, $\alpha=1$, $\sigma=0.2$, and time discretization $\Delta t = 0.1$. As the propagation speed gets larger with $U$, the barycenter of the agent population increases accordingly.}
\label{graph_U_variation}
\end{center}
\end{figure}
\textit{(i)} the individual component $\alpha$ and \textit{(ii)} the interactive component resulting directly from the agent interactions. The extra drift due to mutual interactions lies between 0 (when $U=0$) and  $\frac{\gamma}{2}$ (when $U=\infty$), and it is observed to increase monotonically with $U$. As displayed in Figure \ref{graph_speed_comparison}, the simulated traveling speed obtained  for $U= \infty$ perfectly matches  the  theoretical exact value $\alpha + \frac{\gamma}{2}$.
\begin{figure}[h]
\begin{center}
\hspace{-0.2cm}
\includegraphics[width=140mm, height=50mm]{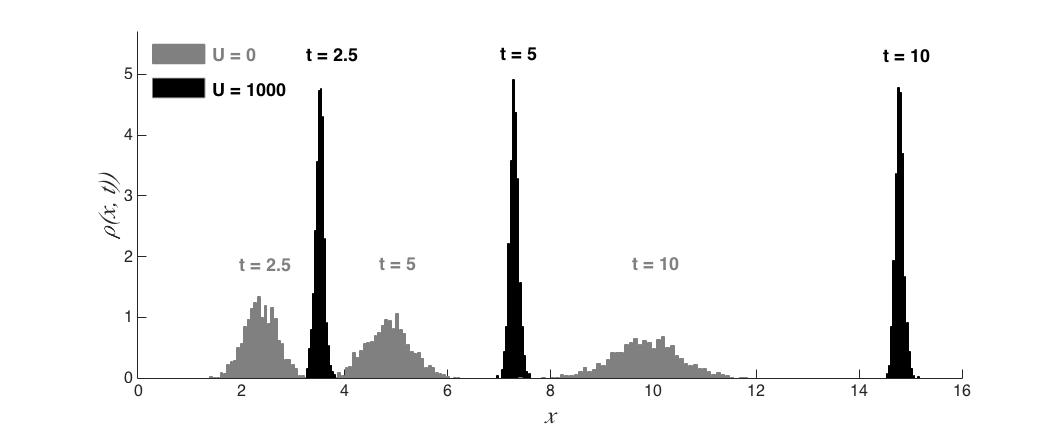}
\caption{Simulated collective dynamics observed for $N=1000$ agents, when $\gamma=1$, $\alpha=1$, $\sigma=0.2$, and time discretization $\Delta t = 0.1$. For $U=1000$, the imitation mechanism generates a collective productivity wave with constant variance, which travels at constant velocity $(\alpha + \gamma/2)$.}
\label{graph_speed_comparison}
\end{center}
\end{figure}
For regimes with $U= \infty$, it is possible to analytically compute only the stationary propagating regime. Nevertheless, as highlighted in Figure \ref{graph_travelling_start},
\begin{figure}[h]
\begin{center}
\hspace{-0.2cm}
\includegraphics[width=140mm, height=50mm]{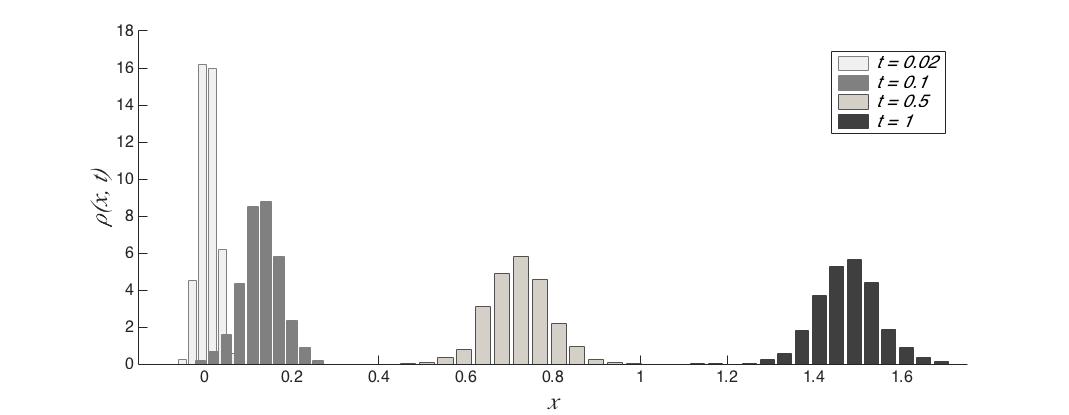}
\caption{Simulated collective dynamics observed for $N=1000$ agents and $U=1000$, when $\gamma=1$, $\alpha=1$, $\sigma=0.2$, and time discretization $\Delta t = 0.1$. After 5 rounds of observation and imitation process between the agents, the collective dynamics reaches its stationary state. }
\label{graph_travelling_start}
\end{center}
\end{figure} 
the simulations clearly show that the transient state before stationarity is reached is definitely very short. Only a few rounds of observation and imitation processes between the agents are necessary to reach the stationary regime predicted by Eq.\eqref{ADENS}.
\end{itemize}

\vspace{0.1cm}
\begin{itemize}
\item[\textit{(c)}] \noindent {\bf Colored Noise Source as Stochastic Driving Sources}\\

\vspace{-0.45cm}
\noindent Strictly speaking, the  WGN can  be only an approximate modeling  of the random environment (the absence of correlations leading to an infinite energy spectrum is obviously never strictly realized). In actual situations, only colored noise processes with finite correlations can be expected. Finite correlations will necessarily introduce memory effects into the dynamics rendering the solutions of the underlying stochastic process non-Markovian. Thus, imposing finite  noise  correlations   enhances  the complexity of the analytic discussion. In \cite{HONGLER2014}, the dynamics of Eq.(\ref{CLIMIT}) when driven by a class of colored noise (namely, the Telegraphic process with exponential correlations similar to the Ornstein-Uhlenbeck process) has been analytically discussed. The discussion in \cite{HONGLER2014} shows that the presence of correlations does not qualitatively alter the set of behaviors unveiled in Section 2.
\end{itemize}

\vspace{0.1cm}
\begin{itemize}
\item[\textit{(d)}] \noindent  {\bf Heterogeneous Populations of Agents}\\

\vspace{-0.45cm}
\noindent In actual populations of agents, heterogeneity may enter in Eq.(\ref{CLIMIT}) in many different ways, including different individual drift functions $f_k(X_k; \vec{X}(t))$, different noise sources $\xi_k(t)$, and obviously different interaction rules ${\cal J}\left(X_k(t), \vec{X}(t) \right)$ (\textit{i.e.} keeping explicit $k$-dependences into Eq.(\ref{CLIMIT})). The observations made in Section 2 are not likely to remain valid for arbitrary heterogeneities. However,  as shown in Figures \ref{graph_burger_total_alpha} and \ref{graph_travelling_total_alpha},
\begin{figure}[h]
\begin{center}
\hspace{-0.2cm}
\includegraphics[width=140mm, height=105mm]{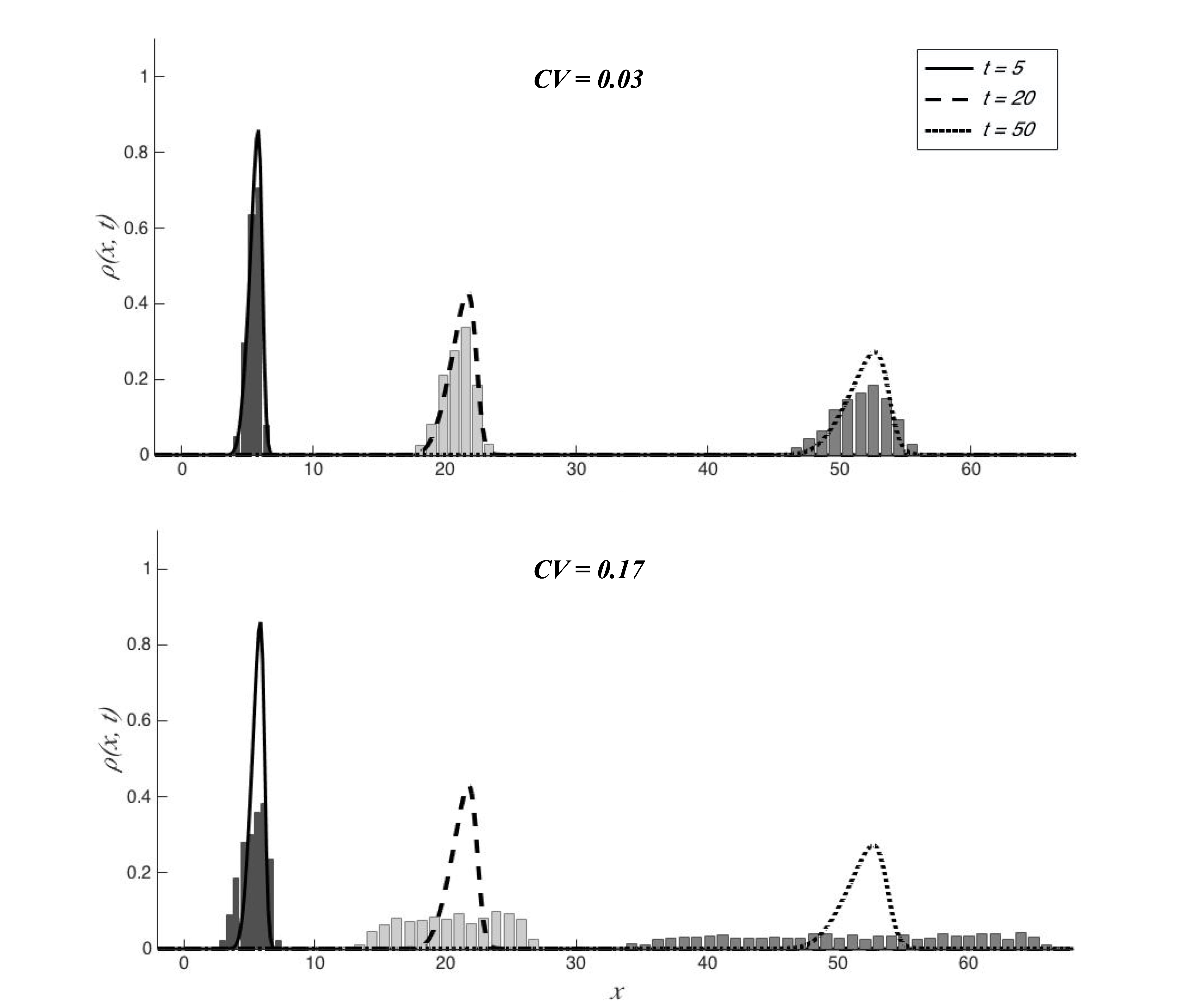}
\caption{Simulated collective dynamics observed for $N=1000$ agents and $U=0.1$, when $\gamma=1$, $\sigma=0.2$, and time discretization $\Delta t = 0.1$. The agents' individual drift $\alpha_k$ is uniformly distributed in $[0.95, 1.05]$ (the top graph, coefficient of variation $CV = 0.03$) and in $[0.7, 1.3]$ (the bottom graph, $CV = 0.17$). The simulated histograms show that the dynamics remains qualitatively robust when the agents' individual drift becomes heterogeneous.}
\label{graph_burger_total_alpha}
\end{center}
\end{figure}
\begin{figure}[h]
\begin{center}
\hspace{-0.2cm}
\includegraphics[width=140mm, height=105mm]{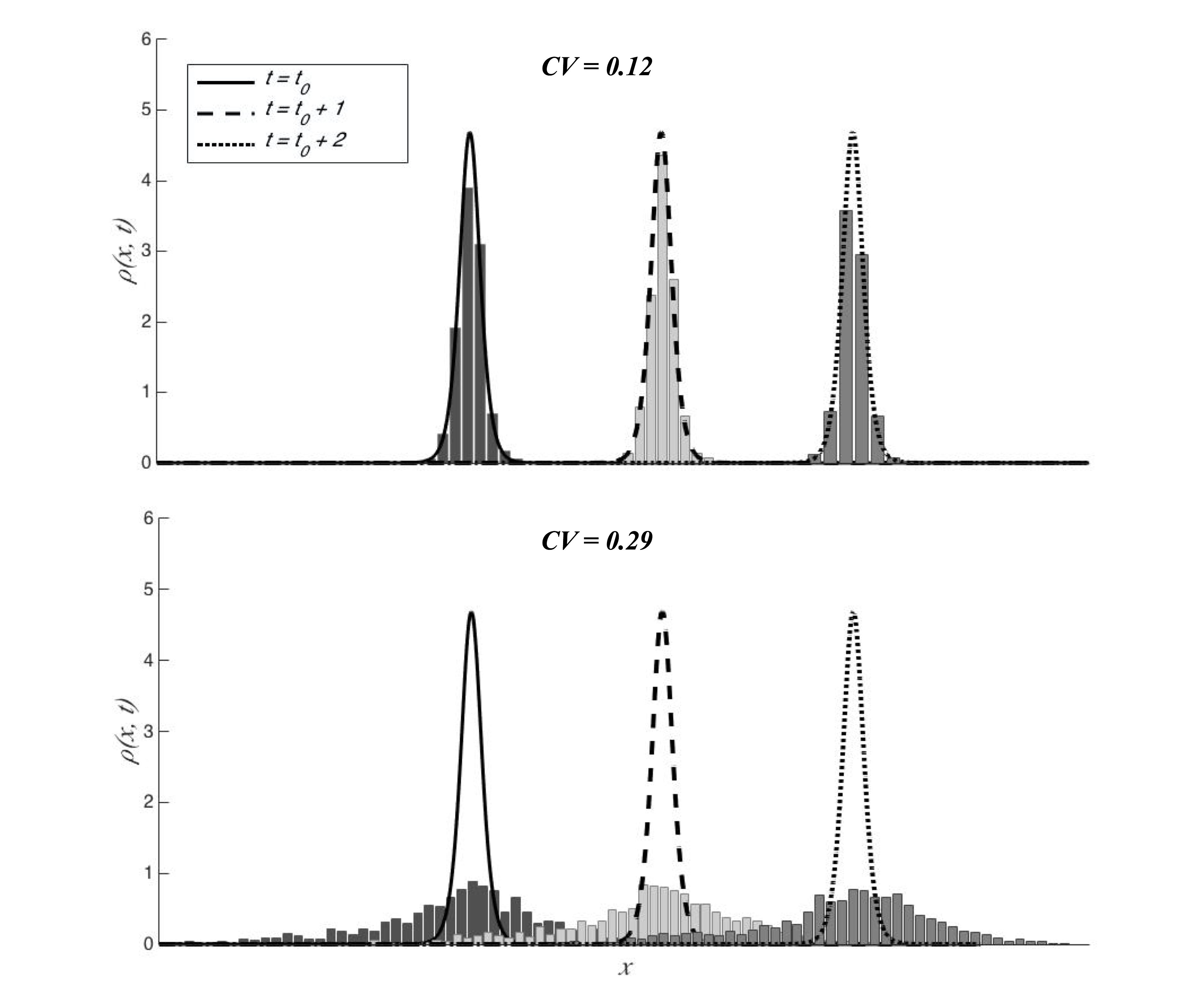}
\caption{Simulated collective dynamics observed for $N=1000$ agents and $U=1000$, when $\gamma=1$, $\sigma=0.2$, and time discretization $\Delta t = 0.1$. The agents' individual drift $\alpha_k$ is uniformly distributed in $[0.8, 1.2]$ (the top graph, coefficient of variation $CV = 0.12$) and in $[0.5, 1.5]$ (the bottom graph, $CV = 0.29$). The simulated histograms show that the dynamics remains qualitatively robust when the agents' individual drift becomes heterogeneous.}
\label{graph_travelling_total_alpha}
\end{center}
\end{figure}
the introduction of heterogeneity in the agents' individual drifts does not qualitatively alter the set of behaviors unveiled in Section 2. Furthermore, for the dynamics expressed in Eq.(\ref{CLIMIT}), heterogeneity may arise from the fact that some agents, belonging to a subset ${\cal E}$, do not obey  the imitation rule implying that ${\cal J}\left(X_k(t), \vec{X}(t) \right)\equiv 0$ for $k \in {\cal E}$. As we observe from Figures \ref{graph_burger_total} and \ref{graph_travelling_total} that the cooperative behavior is qualitatively insensitive to the number $N$ of agents, we can thus safely conclude that the interacting sub-population (\textit{i.e.} the agents who do not belong to ${\cal E}$) continues to exhibit the collective behaviors detailed in Section 2. Other recent analytical discussions for cooperative evolution of specific heterogeneous populations can be found  in \cite{ACEBRON2005} for agents evolving on a circular state space and in \cite{ICHIBA2011} for agents interacting via their ranks.\\

\noindent Table \ref{table_heterogenous_agents} provides a characterization of the theoretical and simulated distributions obtained for the stable growing productivity regime displayed in Figure \ref{graph_travelling_total_alpha}, when the agents'  individual drift fluctuates (but when the mean inclination towards progress remains constant). Simulation experiments do not show a clear tendency on the growth rate in the case of an increase in the variance of the innovation rate of the economical agents. As intuitively expected, the firms' productivity is affected by higher fluctuations when the innovation rate of the firm displays higher variance. Likewise, the asymmetry in the firms' productivity augments with higher differences in the agents' propensity to innovate. This suggests that, in the presence of firms with particularly high innovation rates, imitation mechanisms are not sufficient for laggards to be able to catch the technological frontier. Conversely, simulation experiments do not show a clear trend on the kurtosis of the firms' productivity when their innovation propensity exhibits more fluctuations.

\begin{table}[h!tbp]
\begin{center}
\caption{Characterization of the theoretical and simulated distributions obtained for the stable growing productivity regime displayed in Figure \ref{graph_travelling_total_alpha} for $t=t_0+2$.}
\label{table_heterogenous_agents}
\renewcommand{\tabcolsep}{6pt}
{\scriptsize
\renewcommand{\arraystretch}{1.5}
\vspace{0.3cm}
\begin{tabular}{|c|c|c|c|}
\hline
&$CV=0$ & $CV=0.12$ & $CV=0.29$\\
\hline
\hline
\hline
Mean & 32.9936  & 32.8549 & 32.8762\\
\hline
Variance & 0.0095  & 0.0101 & 0.4661 \\
\hline
Skewness  & 0  & 0.3226 & -0.5575\\
\hline
Kurtosis  & 4.1762  & 5.0038 & 3.4382\\
\hline
\end{tabular}
}
\end{center}
\end{table}
\end{itemize}


\newpage
$\,$
\newpage
\section*{Appendix  F}\label{Hurwitz}

\noindent {\bf Variance of the generalized hyperbolic secant probability laws}. \\

\noindent To fit our  findings with available empirical data (see \cite{KOENIG2016}), we need to calculate the quadrature:
\begin{equation}
\label{QUADRA}
\left\{ 
\begin{array}{l}
{\mathcal I}_{\nu} := {\mathcal I}_{(2-\eta)} = {\cal N}(\nu) \int_{\mathbb{R}} {z^{2} \over \cosh^{2\nu} (z) } dz, \qquad \qquad (2 \nu := 2- \eta >0),\\ \\
{\mathcal N}^{-1}(\nu)  = {4^{\nu} \over 2} {[ \Gamma(\nu)]^{2} \over \Gamma(2 \nu)},
\end{array}
\right.
\end{equation}
\noindent which enters into Eq.(\ref{KMOM1}). To calculate this quadrature, we first use the  moments generating function\footnote{See \cite{GRADSHTEYN1980}, entry 3.512.}:
\begin{equation}
\begin{array}{l} 
{\mathcal R}_{\nu}  (\beta) := \int_{\mathbb{R}} {  \cosh( 2\beta z) \over \cosh^{2\nu}(z)} dz = {4^{\nu}  \over 2}  B \left[ \nu + \beta , \nu-\beta\right] = \\ \\ 
\qquad \qquad   = {4^{\nu} \over 2} {\Gamma(\nu + \beta) \Gamma(\nu- \beta)  \over \Gamma(2\nu) } =  {\mathcal N}^{-1}  (\nu) \left\{ {\Gamma(\nu + \beta) \Gamma(\nu - \beta) \over [\Gamma(\nu)]^{2}}\right\}.
\end{array} 
\end{equation}
\newpage
\noindent Now, we have:
\begin{equation}
\label{GENERF}
\begin{array}{l} 
{d^{2} \over d\beta^{2} } \left[ {\mathcal R}_{2\nu}(\beta) \right] \mid _{\beta =0}=   \int_{\mathbb{R}} {4 z^{2} \cosh (2 \beta z) \over \cosh^{2\nu} (z) } dz \mid_{\beta =0} = \\ \\   \qquad \qquad \qquad \int_{\mathbb{R}} {4 z^{2}\over \cosh^{2\nu} (z) } dz=  {4 {\mathcal I}_{\nu}\over {\mathcal N}(\nu) }.
\end{array} 
\end{equation}

\noindent  Using the notations:
$$
\begin{array}{l}
\Gamma_{\pm } := \Gamma(\nu \pm \beta) \qquad {\rm and } \qquad  \Gamma_{\pm} ^{(k)} = {d^{k}  \over d\beta^{k}} \Gamma(\nu \pm \beta), \\ \\
\Gamma(\nu) ^{(k)} = {d^{k}  \over d\beta^{k}} \Gamma(\nu),
 \end{array}
$$
\noindent and successive derivations with respect to $\beta$ of the right-hand-side of Eq.(\ref{GENERF}), enable us to write:
\begin{equation}
\begin{array}{l}
{d \over d\beta}  {\mathcal R}_{2 \nu}(\beta) =  {{\cal N}^{-1} (\nu) \over [\Gamma(\nu)]^{2}}  \left\{  \Gamma_{+}^{(1)}\Gamma_{-}  - \Gamma_{+}\Gamma_{-}^{(1)}  \right\}  ,\\ \\
{d^{2}  \over d\beta^{2}}  {\mathcal R}_{2 \nu}(\beta) =  {{\cal N}^{-1} (\nu) \over [\Gamma(\nu)]^{2}} \left\{  \Gamma_{+}^{(2)}\Gamma_{-}  - \Gamma_{+}^{(1)}\Gamma_{-}^{(1)}  - \Gamma_{+}^{(1)}\Gamma_{-}^{(1)}  +  \Gamma_{+}\Gamma_{-}^{(2)} \right\}, \\ \\ 
{d^{2}  \over d\beta^{2}}  {\mathcal R}_{2 \nu}(\beta) \mid_{\beta=0} =  2{{\cal N}^{-1} (\nu) \over [\Gamma(\nu)]^{2}}   \left\{ 
 \Gamma^{(2)}(\nu)  \Gamma(\nu) - \left[ \Gamma^{(1)}(\nu)  \right]^{2}\right\} ,
\end{array}
\end{equation}
\noindent implying that:
\begin{equation}
\begin{array}{l}
{d^{2} \over d\beta^{2} } \left[ {\mathcal R}_{2\nu}(\beta) \right] \mid_{\beta=0}= 2 {\mathcal N}^{-1}(\nu)  \left\{ 
{ \Gamma^{(2)}(\nu)  \over \Gamma(\nu) } - \left[ {\Gamma^{(1)}(\nu) \over \Gamma(\nu) }  \right]^{2}\right\} 
 = \\ \\
\qquad \qquad \qquad  2{\mathcal N}^{-1}(\nu) {d^{2} \over d\nu^{2}} \left\{\log [\Gamma(\nu)]\right\}  =2 {\mathcal N}^{-1}(\nu)  \sum_{n=0}^{\infty} {1 \over (n+\nu)^{2}},
\end{array}
\end{equation}
\noindent where we have used the well-known identity:
\begin{equation}
{d^{2} \over dx^{2}} \left\{\log [\Gamma(x)] \right\} \equiv\sum_{n=0}^{\infty} { 1 \over (x+n)^{2}} =  \zeta(2, x), 
\end{equation}
\noindent and where $\zeta (2, \nu)$ is the Hurwitz zeta-function.  Using Eqs.(\ref{QUADRA})  and (\ref{GENERF}) we finally obtain:
\begin{equation}
 {4 {\mathcal I}_{\nu}\over {\mathcal N} (\nu) } = 2 {\mathcal N}^{-1} (\nu)   \sum_{n=0} ^{\infty}  { 1 \over ( \nu +n)^{2}} \quad \Rightarrow \quad {\mathcal I}_{\nu} =  {1 \over 2} \sum_{n=0} ^{\infty}  { 1 \over ( \nu +n)^{2}} := {1 \over 2} \zeta(2, \nu)
\end{equation}
\noindent We can approximate the Hurwitz zeta-function by using:
\begin{equation}
 \zeta(2, \nu) =\sum_{n=0} ^{\infty}  { 1 \over ( \nu +n)^{2}}  \simeq \int_{0}^{\infty} {dx \over (\nu + x)^{2}} = {1 \over \nu} \qquad \Rightarrow \qquad {\mathcal I}_{\nu} \simeq{1 \over 2 \nu}.
\end{equation}


\end{document}